\documentclass[sigconf]{acmart}

\usepackage{algorithm}
\usepackage{algorithmic}
\usepackage{subcaption}
\usepackage{multirow}
\usepackage{bbm}
\usepackage{appendix}

\AtBeginDocument{%
  \providecommand\BibTeX{{%
    \normalfont B\kern-0.5em{\scshape i\kern-0.25em b}\kern-0.8em\TeX}}}

\begin{document}

\title{DMS: Addressing Information Loss with \textit{M}ore \textit{S}teps for Pragmatic Adversarial Attacks}
%\title{DMS: Mind The Information Loss Towards Pragmatic Adversarial Attacks}
\author{Zhiyu Zhu}
\authornote{Corresponding Authors}
\email{nevertough@outlook.com}
\affiliation{%
  \institution{The University of Sydney}
  % \streetaddress{P.O. Box 1212}
  \country{Australia}
}

\author{Jiayu Zhang}
\affiliation{%
  \institution{Suzhou Yierq}
  % \streetaddress{1 Th{\o}rv{\"a}ld Circle}
  % \city{Hekla}
  \country{China}
  }

\author{Xinyi Wang}
\affiliation{%
  \institution{The University of Malaya}
  % \city{Rocquencourt}
  \country{Malaysia}
}

\author{Zhibo Jin}
\affiliation{%
 \institution{The University of Sydney}
 \country{Australia}
 }

\author{Huaming Chen}
\authornotemark[1]
\affiliation{%
  \institution{The University of Sydney}
  \authornotemark[1]
  % \streetaddress{30 Shuangqing Rd}
  % \city{Haidian Qu}
  % \state{Beijing Shi}
  \country{Australia}
  }
\email{huaming.chen@sydney.edu.au}

% \affiliation{%

% \affiliation{%

% \affiliation{%

\renewcommand{\shortauthors}{Zhu, et al.}

\begin{abstract}
Despite the exceptional performance of deep neural networks (DNNs) across different domains, they are vulnerable to adversarial samples, in particular for tasks related to computer vision. Such vulnerability is further influenced by the digital container formats used in computers, where the discrete numerical values are commonly used for storing the pixel values. This paper examines how information loss in file formats impacts the effectiveness of adversarial attacks. Notably, we observe a pronounced hindrance to the adversarial attack performance due to the information loss of the non-integer pixel values. To address this issue, we explore to leverage the gradient information of the attack samples within the model to mitigate the information loss. We introduce the Do More Steps (DMS) algorithm, which hinges on two core techniques: gradient ascent-based \textit{adversarial integerization} (DMS-AI) and integrated gradients-based \textit{attribution selection} (DMS-AS). Our goal is to alleviate such lossy process to retain the attack performance when storing these adversarial samples digitally. In particular, DMS-AI integerizes the non-integer pixel values according to the gradient direction, and DMS-AS selects the non-integer pixels by comparing attribution results. We conduct thorough experiments to assess the effectiveness of our approach, including the implementations of the DMS-AI and DMS-AS on two large-scale datasets with various latest gradient-based attack methods. Our empirical findings conclusively demonstrate the superiority of our proposed DMS-AI and DMS-AS pixel integerization methods over the standardised methods, such as \textit{rounding}, \textit{truncating} and \textit{upper} approaches, in maintaining attack integrity. %Our code is available at:~\href{https://figshare.com/s/047ce3f511a655cfada8}{https://figshare.com/s/047ce3f511a655cfada8}
\end{abstract}




\maketitle

\section{Introduction}
Deep Neural Networks (DNNs) have achieved tremendous success in the field of computer vision~\cite{canziani2016analysis, voulodimos2018deep, buhrmester2021analysis}. In image recognition and classification tasks, DNNs present near-human-level performance~\cite{ciregan2012multi, traore2018deep}. However, DNNs as a core functional module in modern software systems are vulnerable to adversarial examples, which can lead to inaccurate software behaviour~\cite{szegedy2013intriguing}. Therefore, the study of adversarial attacks has become a hot topic in the field of artificial intelligence and software security. Current research suggests that understanding such attacks can further enhance models' robustness against potential threats and improve the overall performance~\cite{goodfellow2014explaining, kurakin2018adversarial, dong2018boosting, madry2017towards, carlini2017towards, inkawhich2019feature, huang2019enhancing}. 

% \begin{figure}[htpb]
%     \centering
%     \includegraphics[width=0.3\linewidth]{images/original.png}
%     \caption{Visualization of model attentions on both the benign image and  DMS-generated adversarial image. The attentions change dramatically on the adversarial image compared with the benign image.}
%     \label{fig:attention1}
% \end{figure}

The core concept of adversarial attacks involves introducing subtle perturbations to images, which are imperceptible to the human eyes but can mislead the target model into making incorrect decisions~\cite{goodfellow2014explaining}. To overcome the practical limitations of traditional white-box adversarial approaches, transferable adversarial attacks have been proposed. The general idea is to generate adversarial examples on a surrogate model, then using them to attack other target unknown models. In this context, the target model is a black box to the attacker, meaning the attacker lacks an understanding of the internal mechanisms of the target model, which aligns more closely with real-world attack scenarios. It implies that the use of transferable adversarial attacks requires performing image processing on the generated data for practical image storage in a specified container format. However, the format used for adversarial examples may result in disparity in attack performance, due to the loss of precision in the traditional way of storing pixel values as discrete numerical entities in image formats.

%In the meantime, the file format in computers is constrained since pixel points are typically stored with discrete values for images. This manifests itself in digital images as spatially discrete points, rather than continuous values. 

% This results in two different types of adversarial attacks, namely adversarial training~\cite{tramer2017ensemble} which generates adversarial samples during training, and adversarial samples~\cite{long2022frequency} which . 

\begin{figure*}[htbp]
  \centering
  % \vspace{-3cm}
  % \includegraphics[scale=0.11]{images/diff_lr.pdf}
  \includegraphics[width=\textwidth]{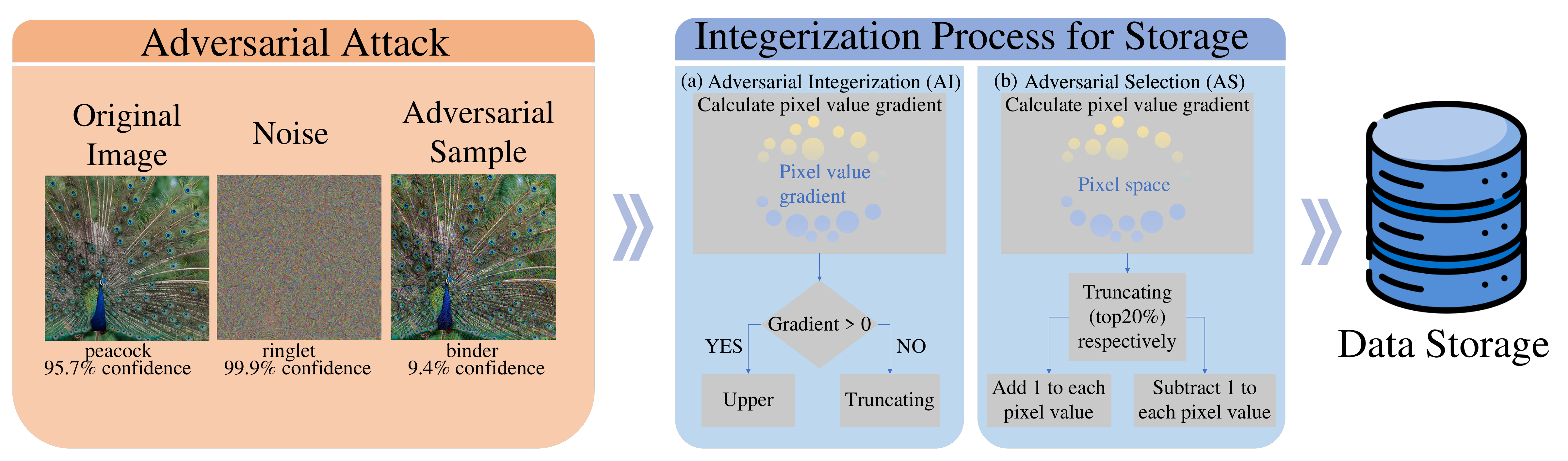}
  \caption{DMS method flow chart (the peacock image is provided as an example using the method from~\cite{goodfellow2014explaining})}
  \label{flow chart}
\end{figure*}
Whether the initial samples undergo manipulation before being stored as adversarial samples, or if a dataset is poisoned with pre-generated adversarial samples prior to an attack, all the adversarial samples stored in computer systems inevitably encounter the problem of the information loss~\cite{wu2020stronger,zhou2021generating}. During image processing, non-integer pixel points are commonly subjected to integerization operations such as directly entering a single digit (\textit{upper approach})~\cite{robidoux2008fast}, rounded by removing digits after the decimal point (\textit{truncating approach})~\cite{hattne2016modeling}, or rounded to the nearest integer (\textit{rounding approach})~\cite{ehsan2015integral}. While the adversarial sample can be effectively generated, the integerization operation performed on the non-integer pixel introduces subtle alterations which impact the image precision.

These subtle changes are almost imperceptible to the human eyes, as it is generally impossible for human to discern minor increases or decreases in pixel values given an image. Since the perturbations added by adversarial attacks are inherently minute, slight fluctuations in pixel values can have a substantial impact on the attack outcome. In particular for transferable attacks. In transferable attacks, adversarial samples are generally saved in image files to attack other target models. Thus, the precision of almost every pixel values in the attack samples changes during the saving process, thereby impacting the effectiveness of the adversarial attack. Ensuring that each non-integer pixel is stored with integerization in a viable direction for adversarial attack can thus be an effective way to mitigate the challenge of precision loss.

% Currently, many adversarial attack methods are based on gradients or their variants. For example, following the pioneering work of FGSM~\cite{goodfellow2014explaining}, numerous gradient-based variants, such as MI-FGSM~\cite{dong2018boosting}, PGD~\cite{madry2017towards}, and SINI-FGSM~\cite{lin2019nesterov}, have emerged. Furthermore, transferable methods in black-box scenarios also rely on gradients. Thus, gradients can be utilised to perform quantization operations. Additionally, our method also takes into account the characteristics of other transferable approaches, being effective against various types of transferable attacks. For instance, generative structures like AdvGAN do not use gradients for attacks but instead use perturbations generated by the generator to approximate the direction obtained from gradients, which does not increase the complexity of the original methods. Moreover, methods similar to transferable attack only require obtaining the gradient in the final step, which does not affect the complexity of the original task.

Currently, many adversarial attack methodologies utilize gradients information. Following FGSM~\cite{goodfellow2014explaining}, a variety of gradient-based methods have been developed, such as MI-FGSM~\cite{dong2018boosting}, PGD~\cite{madry2017towards}, and SINI-FGSM~\cite{lin2019nesterov}. Moreover, transferable approaches in black-box scenarios also rely on gradient information. Notably, generative models such as AdvGAN leverage gradient data during the generator training phase. Backward Pass Differentiable Approximation (BPDA)~\cite{athalye2018obfuscated} effectively circumvents non-differentiable defenses by manipulating gradient during training. This highlights the importance of gradient information in enhancing the adversarial attacks. Therefore, in this work, we investigate the use of gradient information to alleviate the drawbacks introduced by integerization operations.

Building on the considerations mentioned earlier, to address the information loss problem during the storage of adversarial samples, which current methods cannot circumvent, for the first time, we propose a novel method named Do More Steps (DMS), which employs gradient information for integerization operations to mitigate the information loss for adversarial images, shown in Figure~\ref{flow chart}. Specifically, in~Figure\ref{flow chart}(a), we firstly integerize the non-integer pixels based on the gradient direction for discrete image pixels. We name this approach \textit{do more steps via adversarial integerization} (DMS-AI), which seeks to perform an additional adversarial integerization of pixels during image processing to prevent attack failure due to the precision loss.

Moreover, to ensure that \textbf{(i)} changes to pixel values remain infinitesimal when DMS-AI fails, and \textbf{(ii)} to prevent potential issues of gradient oscillation, we draw inspiration from Sundararajan et al.~\cite{sundararajan2017axiomatic} and incorporate the attribution selection method based on integrated gradients in our work. This method is termed \textit{do more steps via attribution selection} (DMS-AS) in Figure~\ref{flow chart}(b), which addresses partial attack failures in DMS-AI and enhances the overall success rate. Specifically, for the DMS-AS step, we increment each original pixel value by one and select a subset of the pixels with the most favorable attribution results for subsequent processing. The augmented pixel values are retained through a truncation operation. Similarly, we decrement each original pixel value by one, then select and truncate a subset of pixels based on attribution results. By segmenting the attributive pixel units and integrating the outcomes of each unit's attribution, detailed in Section~\ref{dmsas}, we aim to mitigate the variance in the loss function landscape caused by non-infinitesimal adjustments. To keep the method simple and effective, we keep the threshold for a subset of the attribution results consistent throughout the whole method. It is important to note that the proportion of selected attribution results, such as 20\%, can be dynamically adjusted. The objective is to maintain a minimal threshold while ensuring a high attack success rate. Additionally, we ensure that the file size of adversarial attack images remains constant before and after applying the DMS algorithm.

% In this paper, DMS algorithm encompasses the optimization strategies of both DMS-AI and DMS-AS. We have thoroughly evaluated the performance of DMS together with using only DMS-AI method and three other conventional techniques. We also investigate the extent of precision loss for each integerization method to demonstrate that the pixel value changes by our method are competently minute. We elaborate this part along with experimental details in Section.~\ref{sec:experiments}.

In this paper, the DMS algorithm encompasses optimization strategies of both DMS-AI and DMS-AS. We have conducted extensive experiments, comparing the performance of DMS with three standardised techniques, across two datasets, ten models, and fourteen attack methods, to provide a thorough evaluation. Additionally, we investigate the precision loss of each integerization approach to quantify the minimal pixel value changes in our method. These aspects, along with detailed experimental procedures, are discussed in Section~\ref{sec:experiments}. The key contributions are as follows:    
\begin{itemize}
    \item We propose DMS to handle information loss in file formats for adversarial attack, and improve the attack success rates.
    \item In line with two different goals, which aim to use gradient ascent for adversarial integerization and integrated gradients for attribution selection, we introduce two different strategies for DMS, namely DMS-AI and DMS-AS. 
    \item A comprehensive statistical analysis is performed based on our benchmarking experiments on different datasets and several adversarial attack methods. The results in Section.~\ref{sec:experiments} outline the state-of-the-art performance of DMS.
    \item We open source the replication package to the community for review and future research.
\end{itemize}

\section{Background and Related Work}
Regarding the attack scenarios where information loss will have a significant impact on attack performance, this section will first introduce various image data format storage methods. Then, we will review conventional non-transferable white-box adversarial approaches. Subsequently, we will introduce the transferable attacks methods. Finally, we will further elucidate the mechanism details using attribution algorithms for consideration.

\subsection{Image Data Format Storage Methods}
In this section, we introduce various image storage formats commonly used in the field of artificial intelligence, including JPEG, PNG, BMP, TIFF, RAW, and OpenEXR.

We begin by discussing those formats that necessitate the use of integers for pixel value storage, which are also the most frequently utilized image data formats in artificial intelligence applications. These include JPEG, PNG, and BMP. The JPEG format employs a lossy compression algorithm for digital photos, typically using 8-bit integers to represent each color channel~\cite{wallace1991jpeg}; PNG is a lossless compression format supporting transparency, storing pixel values in 8 or 16-bit integers~\cite{roelofs1999png}; BMP, an uncompressed image file format, stores pixel values in 24 or 32-bit integers, allocating 8 bits for each color channel~\cite{miano1999compressed}.

On the other hand, there are formats that allow for the storage of pixel values as non-integers, though their application in the field of artificial intelligence is relatively limited and specific to certain domains. These include TIFF, RAW, and OpenEXR. The TIFF format supports 8, 16, or 32-bit integers, as well as 32 or 64-bit floating-point numbers for pixel value storage, making it suitable for applications requiring high color depth or specific data types~\cite{adobe1992tiff}; RAW format captures minimally processed data from digital camera sensors, typically in 12 or 14-bit integers, but high-end cameras may also use 32 or 64-bit floating points~\cite{andrews2006raw}; OpenEXR is specifically designed for high dynamic range (HDR) images, capable of storing pixel values as half or full precision floating points, suitable for complex lighting and color scenes~\cite{kainz2003openexr}.

Due to constraints in computational resources and training efficiency, the majority of image storage formats in the field of artificial intelligence predominantly utilize integer storage formats, as opposed to floating-point formats. Consequently, our focus is primarily directed towards the former.

\subsection{White-box Adversarial Attacks}
% Compare IG with SM
%Computation complexity and efficiency
Gradient-based methods have been widely used in adversarial attacks. Goodfellow et al.~\cite{goodfellow2014explaining} proposed the Fast Gradient Sign Method (FGSM) algorithm, which utilises gradient information for generating adversarial samples. It is a target-free attack method (i.e., the adversarial samples are not specified as a particular class, as long as they are different from the original class). Unlike using gradient descent to optimize the loss function to find the minimum, FGSM uses a one-step gradient ascent to train the attack samples. The Iterative Fast Gradient Sign Method (I-FGSM) is an advanced version proposed by Kurakin et al.~\cite{kurakin2018adversarial}. I-FGSM iteratively adds a certain amount of perturbation to the original sample, causing the classifier to make errors in prediction. Compared to FGSM, I-FGSM is able to obtain better attack results. 

The projected gradient descent (PGD) algorithm is proposed as an adversarial sample generation algorithm by Madry et al.~\cite{madry2017towards}. Different from FGSM and I-FGSM which only perform one gradient update, PGD generates deceptive adversarial samples by searching for the optimal solution by multiple iterations. During the iterative process, the algorithm projects perturbations into the prescribed range with each small step, directing each iteration towards the optimal direction. Also, during the iteration, the size of the adversarial perturbation is limited using $L_{p}$ parametrization, which controls the visual perceptual difference of the adversarial samples. PGD performs better in terms of attack effectiveness and stability compared to other adversarial sample generation algorithms, particularly against defense mechanism. In addition, the PGD algorithm allows for the step size and the number of iterations adjustment to further improve the attack efficiency.

Unlike classic gradient-based methods to search for gradient update for perturbation, Carini and Wagner attack (C\&W)~\cite{carlini2017towards} designs a loss function to measure the difference between input and output. The loss function contains adjustable hyperparameters to calculate the confidence of the generated adversarial samples. The goal is to minimize the distance between the input data and the target classification while maximizing the distance between the adversarial samples and other classes to attack the model. In addition, the C\&W algorithm can control the confidence of the generated adversarial samples to prevent misclassification. By reasonably adjusting the hyperparameters and confidence parameters in the algorithm, it is possible to generate adversarial samples with high quality, high confidence, and high attack success rate.

\subsection{Transferable Adversarial Attacks}

This section focuses on introducing different types of transferable attack methods, which, based on their underlying principles, can be divided into four categories. Firstly, the gradient editing method involves generating attack samples by modifying or optimizing gradient information. Secondly, the construction of semantic similarity emphasizes identifying samples that are semantically related to the target, thereby enhancing transferability. Thirdly, the modification of attack targets exploits the similarities between different models to facilitate effective cross-model attacks. Lastly, the generative structure approach employs generative adversarial networks to emulate the decision boundaries of the target attack model, effectively generating attack samples with strong transferability.

% \subsubsection{Gradient-Based Methods in Black-Box Adversarial Attacks}

\subsubsection{Gradient Editing}
Considering problems such as local maxima in sample optimization, Dong et al. optimized the I-FGSM basis and proposed the Momentum Iterative Fast Gradient Sign Method (MI-FGSM)~\cite{dong2018boosting}. The principle of this method involves the introduction of a momentum term, which enhances the stability and convergence speed during the optimization process, allowing the optimization to escape local optima. Overall, MI-FGSM achieves higher attack efficacy for most deep learning models, demonstrating a degree of universality and portability.

Gradient Relevance Attack (GRA)~\cite{zhu2023boosting} incorporates two novel strategies: a gradient relevance framework, which utilizes neighborhood information to adaptively correct the direction of updates during the generation of adversarial examples, and a decay indicator that modulates the step size to mitigate fluctuations in the sign of adversarial perturbations. These innovations are aimed at exploiting the observed fluctuation phenomena within adversarial perturbations to enhance the transferability of generating adversarial examples.

\subsubsection{Construction of Semantic Similarity}
Diverse Input Fast Gradient Sign Method (DI-FGSM)~\cite{xie2019improving} enhances the transferability of adversarial examples through the incorporation of input diversity. Its fundamental concept involves embedding input diversity during the adversarial sample generation process to identify Semantic Similarity. This goal is realized by executing stochastic modifications (like resizing and padding) to the input image in every iteration. Such alterations result in a spectrum of diverse input configurations, which aids in averting overfitting to particular network parameters and thus amplifies the efficiency of the adversarial examples across various models.

Dong et al.~\cite{dong2019evading} introduce a Translation-Invariant Fast Gradient Sign Method (TI-FGSM) for generating more transferable adversarial samples against defense models. TI-FGSM optimizes perturbations over a set of translated images to create adversarial samples that are less sensitive to the white-box models being attacked, thereby enhancing the transferability. Moreover, to enhance the attack efficiency, \cite{dong2019evading} demonstrates that convolving the gradients for non-translated images with predefined kernels can be effective. TI-FGSM is applicable to any gradient-based attack approach.

Lin et al.~\cite{lin2019nesterov} presents the Scale-Invariant Nesterov Iterative Fast Gradient Sign Method (SINI-FGSM), which combines the Nesterov accelerated gradient and scale invariance principles. The primary aim is to enhance the efficiency of generating adversarial samples and improve their transferability. SINI-FGSM employs the Nesterov accelerated gradient for updating perturbations and integrates a scale-invariant attack strategy to target images of varying scales. The essence of this algorithm lies in the simultaneous consideration of gradient optimization and scale invariance during the perturbation optimization process, resulting in the generation of more efficient and highly transferable adversarial samples.

It is noteworthy that traditional gradient-based methods primarily focus on spatial domain perturbations. However, these methods have demonstrated some limitations when dealing with subtle variations in behavior between the source and target models. In such cases, they may fail to capture the intricate relationships among image pixels comprehensively. To address this concern, Long et al.~\cite{long2022frequency} introduce the Spectrum Simulation Attack (SSA) attack method, which delves into the impact of gradients on the transferability of adversarial attacks in the frequency domain.

\subsubsection{Modification of Attack Targets}
% Unlike traditional gradient-based attack methods, there exists a category of attack methods primarily focusing on key features that models pay attention to. This approach is based on the hypothesis that features of interest across different models exhibit similarities, thereby enhancing the transferability of generated adversarial examples.
The Feature Importance-aware Attack (FIA)~\cite{wang2021feature} method exemplifies a feature-based adversarial attack strategy. It aims to enhance the transferability of adversarial examples by targeting crucial, object-aware features that significantly impact decisions across various model architectures. FIA employs an aggregate gradient method, averaging gradients from randomly transformed versions of an input image. This process emphasizes consistent and object-related features, reducing the influence of model-specific "noisy" features. By focusing on these critical features, FIA can create adversarial examples that are more likely to be effective across different models, thus improving the transferability of adversarial samples.

The Momentum Integrated Gradients (MIG)~\cite{ma2023transferable} primarily characterize its adversarial examples as effective against both Vision Transformers (ViTs) and Convolutional Neural Networks (CNNs). Its core principle involves using integrated gradients, as opposed to regular gradients, to guide the creation of adversarial perturbations. This method is predicated on the observation that integrated gradients exhibit higher similarity across different models compared to regular gradients. MIG also implements a momentum-based iterative strategy, accumulating gradients from previous iterations with current ones, thereby iteratively modifying the perturbations. This approach not only enhances the attack success rate but also improves the transferability of adversarial examples across different architectural models.

The Neuron Attribution-based Attack (NAA)~\cite{zhang2022improving} method fundamentally strengthens feature-level attacks by providing a more precise estimation of neuron importance in deep neural networks (DNNs). This technique involves attributing the output of a DNN model to individual neurons in an intermediate layer, followed by an approximation scheme to reduce computational overhead. This method allows for weighting neurons based on their attribution, leading to more effective and efficient feature-level attacks. It also leverages the potential similarity in features across models to enhance the transferability of generated samples.

The Double Adversarial Neuron Attribution Attack (DANAA)~\cite{jin2023danaa}, similar to NAA, uses mid-layer features of models to increase the transferability of adversarial samples. However, DANAA diverges from NAA by adopting a non-linear gradient update path for neuron attribution. This non-linear approach effectively circumvents local optima in attribution, offering a more detailed and precise understanding of neuron contributions. This enhances the accuracy of importance attribution across different neurons. Therefore, adversarial examples generated by DANAA are not only more effective in their target model but also demonstrate improved transferability.

The Structure Invariant Attack (SIA)~\cite{wang2023structure} technique operates on the premise of executing a multitude of stochastic alterations on an image, to generate a varied array of adversarial examples that retain structural attributes. In the SIA approach, the image is segmented into blocks, each undergoing stochastic transformations like rotation and scaling, thus augmenting the variety of the samples and facilitating the identification of analogous semantics. While preserving the essential framework of the original image, this technique generates challenging adversarial examples that are adept at misleading deep neural networks. 

\subsubsection{Generative Structure}

Moreover, a distinct category of transferable attack methods, grounded in the principles of Generative Adversarial Networks (GANs), has emerged. Within this innovative field, the Adversarial Generative Adversarial Network (AdvGAN)~\cite{xiao2018generating} stands as the pioneering method, being the first to be proposed in this category. It is meticulously designed to craft adversarial examples capable of misleading deep neural networks. The architecture of AdvGAN is composed of three integral components: a generator, a discriminator, and the targeted neural network. The generator's function is to create perturbations derived from the original instance, whereas the discriminator ensures that these generated instances are indistinguishable from legitimate data. AdvGAN employs a sophisticated loss function that integrates adversarial loss, hinge loss, and GAN loss. Once adeptly trained, AdvGAN is capable of efficiently generating adversarial perturbations for any given input instance.

GE-AdvGAN~\cite{zhu2024geadvgan} is also a transferable attack method based on GANs. Its core concept involves optimizing the training process of generator parameters in Adversarial Generative Models (AGMs). A key aspect of GE-AdvGAN is the incorporation of a Gradient Editing (GE) mechanism, which, in conjunction with frequency domain exploration, aids in determining the direction of gradient editing. This enhances the efficacy of adversarial samples and increases the efficiency of the algorithm.

The Frequency-based Stationary Point Search (FSPS)~\cite{zhu2023improving} algorithm uses semantically similar samples to enhance the transferability of adversarial attacks. It aims to enhance the transferability of adversarial attacks in black-box scenarios by leveraging stationary points on the loss curve and conducting a frequency search to determine the most effective direction for an attack. The key principle of FSPS is to locate a stationary point, which is a point where the first derivative of a function is zero, on the loss curve of the DNN. This stationary point serves as a starting point for the subsequent search for local optimal adversarial points. Additionally, FSPS employs frequency domain analysis to determine the direction of the attack from these stationary points. This strategy involves transforming image space into a frequency domain using Discrete Cosine Transformation (DCT) and then applying perturbations to identify sensitive frequencies for the attack. 

% While those adversarial attacks may refer to either adversarial training or adversarial sample generation towards targeted black-box models, they all inevitably encounter the problem of storage precision loss. Therefore, in this work, we target the gradient-based adversarial attack algorithms to analyse the impacts of storage precision loss.
While adversarial attacks may pertain to either adversarial training or adversarial sample generation targeting black-box models, they inevitably face the issue of storage precision loss. Hence, in this work, we focus on gradient-based adversarial attack algorithms to analyze the impacts of storage precision loss. The results validate that our method is effective across the four mentioned transferability approaches, demonstrating its universality.

%Depending on the different attacking needs, which can either refer to adversarial training or to generating and saving adversarial samples on a source model first, and subsequently launching attacks on the target black-box model, these gradient-based adversarial attack algorithms will inevitably encounter the problem of storage precision loss. At the same time, white-box attacks have the great advantage of studying performance such as model robustness. Black-box attacks are more aligned with adversarial attacks in real-world scenarios, as obtaining model information is often not easily achievable in practical environments. 
%Therefore, it is imperative to study the storage precision loss of gradient-based adversarial attack algorithms.

\subsection{Attribution Methods}
Since deep neural network architectures exhibit nonlinear relationships between the input and out, we herein consider the attribution methods, particularly the gradient-based attribution methods, to interpret the model outcome. Some exemplar methods include Integrated Gradients (IG)~\cite{sundararajan2017axiomatic}, Adversarial Gradient Integration (AGI)~\cite{pan2021explaining}, and Guided Integrated Gradients (GIG)~\cite{kapishnikov2021guided}. 

The Integrated Gradients method is more computationally efficient compared to the Saliency Map (SM) method~\cite{simonyan2013deep,patra2020incremental}. However, on the downside, IG is a model-specific method and therefore requires knowledge of the internal structure of the model to calculate the gradients of the model. Thus, Pan et al.~\cite{pan2021explaining} combine IG with adversarial attacks as the Adversarial Gradient Integration (AGI) method to integrate the gradients from adversarial samples. AGI explores the model's robustness by adding adversarial perturbations. In this way, a more accurate attribution analysis will be obtained by evaluating the sensitivity of model prediction to each input feature. The advantage of AGI is that it does not require reference points. However, in the case of high-dimensional input space and complex nonlinear functions, AGI could not perform well. Based on the findings, Kapishnikov et al.~\cite{kapishnikov2021guided} propose GIG to further leverage guided gradients to eliminate unnecessary noisy pixel attribution. GIG constrains the network input and then back-propagates the gradient of neurons so that only the pixel attributes relevant to the predicted category will be retained. It effectively reduces the negative effect of noise generated in irrelevant regions on the attribution results. 

% While there are different attribution methods, we focus on IG algorithm in this work as the primary method for the DMS attribution selection step. Another reason is that IG requires much less computational time thus can guarantee the efficiency of DMS method. However, we consider that AGI, GIG and other attribution methods can be easily incorporated in our algorithm. 
While various attribution methods exist, in this work, we primarily focus on the Integrated Gradients (IG) algorithm as the key method for the DMS attribution selection step. There are two main reasons for this choice. Firstly, IG provides a clear baseline for rounding pixel values up or down, which we discuss in detail in Section~\ref{details DMS-AS}. Another reason is that IG requires significantly less computational time, thereby ensuring the efficiency of the DMS method. Additionally, we consider that other attribution methods, such as AGI and GIG, can be easily incorporated into our algorithm.

\section{Method} 
\subsection{Problem Definition}
Given the RGB pixels of the adversarial sample before saving in the file, we denote it as $P=(a,b,c)$. Due to the discrete nature of pixels, the non-integer value $P_{non}=(a',b',c')$ is generally integerized, then the non-integer pixel value $P_{non}$ will be transformed into the integer form $P'=([a'],[b'],[c'])$. In such cases, the variation of pixel values resulting from integerization will cause information loss for storage precision, which in turn affects the adversarial attack success rate. Usually, non-integer pixel points will be directly entered into either one digit (upper), rounded by removing digits after the decimal point (truncating), or rounded to the nearest integer (rounding). Hence, determining the most effective aproach to minimise the impact of precision loss while integerizing each pixel point for enhanced attack performance poses a challenge.

Assuming we have a deep neural network $N:R^{n}\to R^{m}$ and the original image sample $x^{0}\in R^{n}$, which original label is $s$. The adversarial sample optimized by the attack algorithm is $x$, whose adversarial label is $t$. In this paper, we are committed to exploring an approach to ensure non-integer pixel point integerizing in a more meaningful direction for attacks. In other words, we will ensure that after the integerizing operation, the stored sample $x'$ can still satisfy the following equation:
\begin{equation}
     \left \| x'-x^{0} \right \| _{n}<\epsilon  \quad subject \ to\quad N(x')\ne N(x^{0})
\end{equation}
Where $\left \| \cdot \right \|_{n} $ represents the n-norm distance. The sample $x'$ can still be classified as the adversarial label $t$ by the deep neural network $N(x)$, meaning that the adversarial sample has been successfully retained in the file. To achieve this goal, our method is divided into two parts: adversarial integerization and attribution selection.
\begin{itemize}
\item[·] \textit{Adversarial integerization}  
\\ Integerize non-integer pixel points by gradient $\bigtriangledown_{x} J(x,t)$ to reduce attack failure due to loss of storage precision.
\item[·] \textit{Attribution selection}
\\ Select 20\% pixels with the best attribution results upon value adjustment (plus one or minus one) respectively, and these pixels will be further truncated. This step aims to resolve the potential attack failure that may still exist after \textit{Adversarial integerization}.
\end{itemize}
\subsection{DMS for Adversarial Integerization}
\subsubsection{Motivation of DMS-AI}
Updating the samples in the gradient descent direction to maximize the loss function, which increases the probability that the model classifier discriminates a misclassification, has been an effective means of adversarial attacks. We note that the non-integer pixel integerization process when storing images can use the gradient of adversarial samples to ensure that the integerized samples can still easily attack the target model. Thus, the uncertain non-integer pixel value storage can be transformed into a more target-oriented storage for adversarial attacks.
\subsubsection{Details of DMS-AI}
\begin{figure}[htbp]
  \centering
  % \vspace{-3cm}
  % \includegraphics[scale=0.11]{images/diff_lr.pdf}
  \includegraphics[width=\linewidth]{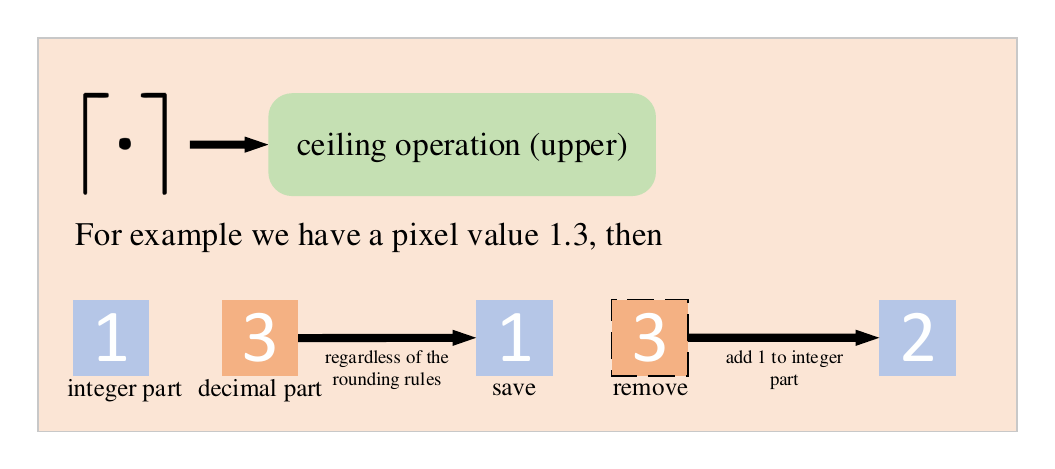}
  \caption{Schematic diagram of ceiling operation}
  \label{fig:ceiling}
\end{figure}

\begin{figure}[htbp]
  \centering
  % \vspace{-3cm}
  % \includegraphics[scale=0.11]{images/diff_lr.pdf}
  \includegraphics[width=\linewidth]{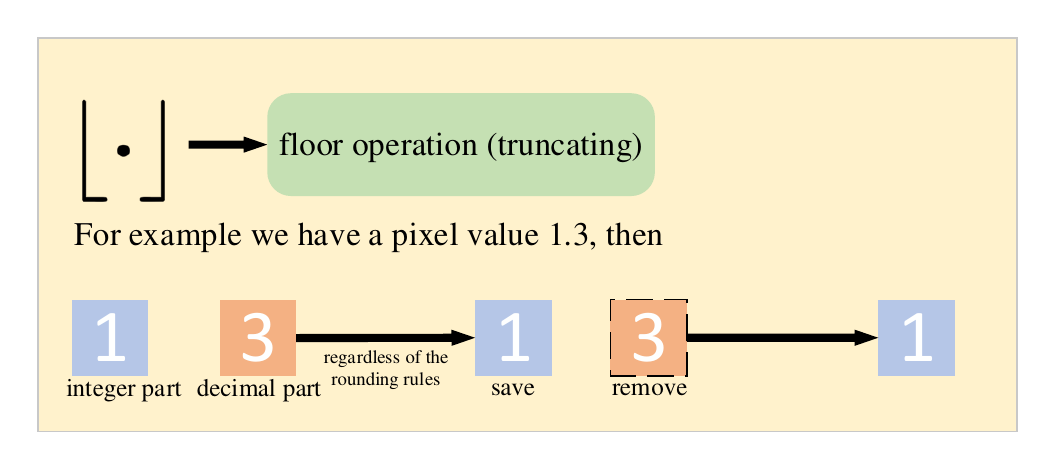}
  \caption{Schematic diagram of floor operation}
  \label{fig:floor}
\end{figure}

Suppose we have a set of pixel values $Q=\left \{ P_1,P_2,...,P_{W \times H} \right \}$. Let $p=(a,b,c)$ denote the RGB pixel values in adversarial sample $x$ (with $W \times H$ pixels), where the pixel values $a,b,c$ are integers or non-integers. We use the gradient $\bigtriangledown_{x} J(x,t)$ to determine how $P'$ is integerized. Since there are $W \times H $ pixel values in the set $Q$, $W \times H \times 3$ gradients are required. We define $\bigtriangledown_{x(\theta )} J(x(\theta ),t)$ as the gradient of the pixel value $\theta$. In DMS-AI, we define the integerization method as $\odot [\cdot]$, specifically as follows:
\begin{equation}\label{DMS-AI-1}
    \odot[\theta]=\left \lceil \theta \right \rceil  \quad s.t.\quad \bigtriangledown_{x(\theta)} J(x(\theta),t)>0 
\end{equation}
\begin{equation}\label{DMS-AI-2}
    \odot[\theta]=\left \lfloor \theta \right \rfloor  \quad s.t.\quad \bigtriangledown_{x(\theta)} J(x(\theta),t)<0
\end{equation}
Where $\left \lceil \cdot \right \rceil$ represents ceiling operation, that is, regardless of the rounding rules, it is ignored as long as the pixel value has a decimal part. The integer par will be added with one as shown in Figure.\ref{fig:ceiling}. On the contrary, $\left \lfloor \cdot \right \rfloor$ represents floor operation, that is, regardless of the rounding rules, as long as the pixel value has a decimal part, ignore it (as shown in Figure.\ref{fig:floor}).

Then for $P = (a, b, c)$, we can obtain the integerized $P'$
\begin{equation}
    P'=(\odot[a],\odot[b],\odot[c])
\end{equation}
We further perform ceiling and floor operations on the pixel values in $Q$ when the gradient value is positive and negative, respectively, so that the corresponding values of gradient ascent that are meaningful to the attack can be rounded up and stored. The gradient descent that is meaningless to the attack will be rounded down.
\subsection{DMS for Attribution Selection}\label{dmsas}
\subsubsection{Motivation of DMS-AS}
For the attribution selection step, we aim to further bolster the DMS performance once the attack fails via the DMS-AI step. Additionally, the latent oscillation problem in gradient ascent is an issue for DMS. Thus, we incorporate the Integrated Gradient method into the DMS algorithm as the DMS-AS step, which can further ensure minimal precision changes in image processing. Here IG serves as the baseline for pixel importance evaluation, since the process of DMS-AS needs to evaluate which pixels most significantly influence the adversarial attack when all pixels are either +1 or -1.
\subsubsection{Details of DMS-AS} \label{details DMS-AS}
Formally, we assume a deep neural network $N(\cdot)$ where $N(x)$ represents the output of the input image $x$. Specifically, $x^{0}\in R^{n}$ is the baseline input, which could be a black image. We let the gradient accumulate along a linear path from the baseline $x^{0}$ to the input $x$. The outcome from each pixel point is the attribution value we will describe in detail below.

Given the gradient of input image $x$ (with $W \times H \times 3$ pixels) to be $\frac{\partial N}{\partial x_{i}} $, we define the attribution results with respect to the baseline  $x^{0}$ as
\begin{equation}\label{eq4}
    A:=\sum_{i=1}^{W \times H \times 3} (x_{i}-x^{0}_{i})\int_{0}^{1} \frac{\partial N}{\partial x_{i}}(x^{0}+\alpha (x-x^{0})) \mathrm{d}\alpha
\end{equation}

Where $W$ represents the width of the image, $H$ represents the height of the image, and $c$ represents the number of channels.
%修改 只需要说明A代表队从baseline变成当前样本的attribution
In Eq.~\ref{eq4}, $(x^{0}+\alpha (x-x^{0}))$ represents a linear path, along which the gradient $\frac{\partial N}{\partial x_{i}} $ will be accumulated. According to the fundamental theorem of calculus, $A$ represents the attribution integrated along the baseline input into the current input.

It is clear that integration paths can be linear or nonlinear. In this section we define our linear path $(x^{0}+\alpha (x-x^{0}))=\delta(\alpha)$ for $\alpha \in [0,1]$. Then for baseline input $x^{0}$, $\delta(0)=x^{0}$. Similarly when $\alpha = 1$, $\delta(1)=x$. The attribution with respect to the path function $\delta(\alpha)$ is defined as
% \begin{equation}\label{eq5}
%     A_{\delta}:=\sum_{i=1}^{N^{2}} (x_{i}-x^{0}_{i})\int_{0}^{1} \frac{\partial N(\delta(\alpha))}{\partial \delta_{i}(\alpha)}\frac{\partial \delta_{i}(\alpha)}{\partial \alpha}\mathrm{d}\alpha 
% \end{equation}

\begin{equation}\label{eq5}
    A_{\delta}:=\sum_{i=1}^{W \times H \times 3} (x_{i}-x^{0}_{i})\int_{0}^{1} \frac{\partial N(\delta(\alpha))}{\partial x_{i}}\mathrm{d}\alpha 
\end{equation}

To calculate the specific integration path of $\delta(\alpha)$ from $x^{0}$ to $x$, based on the Riemann sum theorem, we divide $m$ pixel units along the line. Finally, Eq.~\ref{eq5} can be expressed as
% \begin{equation}
%     A_{\delta}:= \frac{1}{m} \sum_{k=1}^{m}(\frac{\partial N(\delta(k))}{\partial \delta_{i}(k)}) (\sum_{i=1}^{N^{2}}(x_{i}-x^{0}_{i})\frac{\partial \delta_{i}(k)}{\partial x_{i}} )    
% \end{equation}
\begin{equation}\label{eq:AS}
    A_{\delta_{i}}:\approx \frac{1}{m} \sum_{k=1}^{m}((x_{i}-x^{0}_{i})\frac{\partial (\delta(k))}{\partial x_{i}} )    
\end{equation}
% two ways to update the baseline, +1 or -1
Where $\delta(k)=x^{0}+\frac{k}{m}(x-x^{0})$ are the divided pixel units, we have the attribution result corresponding to the $i$-th pixel value. Following that, we can select non-integer pixel values with the best attribution results, ensuring that the overall attack success rate can be retained despite the attack may initially fail after the DMS-AI step. It is worth noting that we have two baseline operations for the pixel values after the DMS-AI step, which are addition by one and subtraction by one. We select 20\% of the best attribution results of each pixel value respectively every time we add or subtract 1 to the pixel value domain. The number of selected attribution results (i.e., 20\%) can be dynamically adapted depending on the final attack performance.

\begin{algorithm}
	%\textsl{}\setstretch{1.8}
	\renewcommand{\algorithmicrequire}{\textbf{Input:}}
	\renewcommand{\algorithmicensure}{\textbf{Output:}}
	\caption{DMS-AI}
	\label{alg:DMS-AI}
	\begin{algorithmic}[1]
	\REQUIRE Model $m$, Input data $x$, perturbation $\epsilon$, Attack steps $s$, Attack method $Ma$
    \ENSURE $x^{*}$
    \STATE $x^\prime = M_a(x,m,\epsilon,s)$
    \STATE Calculate the loss function using $x^\prime$ forward propagation
    \STATE Backpropagate the loss function and get the gradient $x^\prime.grad$
    \STATE Update $x^\prime$ by using $x^\prime.grad$
    \STATE Get $x^{*}$ by Equation~\ref{DMS-AI-1} and Equation~\ref{DMS-AI-2}
\RETURN $x^{*}$
	\end{algorithmic}  
\end{algorithm}

\begin{algorithm}
	%\textsl{}\setstretch{1.8}
	\renewcommand{\algorithmicrequire}{\textbf{Input:}}
	\renewcommand{\algorithmicensure}{\textbf{Output:}}
	\caption{DMS-AS}
	\label{alg:DMS-AS}
	\begin{algorithmic}[1]
	\REQUIRE  Ratio of Operation $k$, Attack sample after DMS-AI $x$, Attack method $M_a$
    \ENSURE $x^{*}$
    \STATE $x^0 = \operatorname{clip}(x+\mathbbm{1} ,0,255)$  \quad \texttt{\# $\mathbbm{1}  \in  R^{W \times H \times 3}$}
    \STATE Calculate $A_{\delta_{i}}$ by Equation~\ref{eq:AS}
    \FOR {$i$ in $I(Top_k (A_{\delta_{i}}))$}
        \STATE $x_i = x_i + 1 $
    \ENDFOR
    \STATE $x^0 = \operatorname{clip}(x-\mathbbm{1} ,0,255)$ 
    \STATE Calculate $A_{\delta_{i}}$ by Equation~\ref{eq:AS}
    \FOR{$i$ in $I(Top_k (A_{\delta_{i}}))$}
        \STATE $x_i = x_i -1 $
    \ENDFOR
    \STATE $x^{*} = \operatorname{clip}(x_i,0,255)$
\RETURN $x^{*}$
	\end{algorithmic}  
\end{algorithm}

As shown in the \textbf{Algorithm~\ref{alg:DMS-AI}}, the process of the DMS-AI algorithm begins with the generation of an initial adversarial sample using an attack method, characterized by input data, perturbation parameters, and a predefined number of attack steps. Subsequently, the algorithm computes the loss function through the forward propagation of this sample. Crucial to the approach is the backpropagation of this loss function to obtain the gradient of the adversarial sample, which is then utilized to update the sample itself. The final and pivotal step involves refining the adversarial sample by applying specific equations, aimed at mitigating the precision loss effect during storage. The output, a refined adversarial sample, is expected to demonstrate enhanced effectiveness due to its reduced susceptibility to precision loss.

As shown in \textbf{Algorithm~\ref{alg:DMS-AS}}, the DMS-AS algorithm, an extension of the DMS-AI method, aims to further refine adversarial samples by employing a baseline adjustment and an attribution scoring mechanism. The symbol $\mathbbm{1}$ in line 1 and line 6 represents a matrix of dimensions W $\times$ H $\times$ 3, where all values are set to 1. $I$ in line 3 and line 8 refers to obtaining the index of the matrix. Initially, the algorithm creates a baseline $x^0$ for the adversarial sample $x$ processed by DMS-AI, using a clip operation to ensure the sample values remain within the standard 0 to 255 range. This is achieved by adding or subtracting a unit matrix, which matches the dimensions of the sample. The reason of not using a black image in IG as the baseline is that black images result in a greater loos of semantic information compared to our baseline. Since DMS-AS is the remedy for DMS-AI, its purpose is to obtain the pixel importance of sample $x$ processed by DMS-AI after changing one pixel. Utilizing a black image as the baseline would significantly dimish the accuracy of attribution. The core of the algorithm involves computing the $A_{\delta_{i}}$ using Equation~\ref{eq:AS}, followed by identifying and modifying the $Top_k$ elements based on these scores. This process involves incrementing or decrementing these key elements of $x$, thereby fine-tuning the adversarial characteristics of the sample. Consequently, DMS-AS enhances the adversarial sample's potency. Overall, the entire DMS process solves the problem of information loss during sample storage and has little impacts on sample stealthiness, since the AS operation is only activated in rare cases where AI operation fails.

\section{Evaluation}\label{sec:experiments}
\subsection{Datasets}
% Follow the setup of A,B,C We followed the experiment setting of several adversarial attack method
Following the dataset generation methods referenced in Zhang et al.~\cite{zhang2022improving} and Wang et al.~\cite{wang2021feature}, we randomly sample 1000 images of different categories from the CIFAR-100~\cite{krizhevsky2009learning} and ImageNet~\cite{krizhevsky2017imagenet} datasets as our experimental dataset. In white-box performance testing, we utilized both CIFAR-100 and ImageNet datasets. For black-box transferability experiments, we adhere to the general practice of other black-box algorithms by employing the ImageNet dataset for testing.

\subsection{Models}
In our research, we employed a total of ten models, which can be categorized into two groups: seven models trained using conventional methods and three models that underwent defensive training. The conventionally trained models include: Inception-v3 ~\cite{szegedy2016rethinking}, Inception-v4, ResNet-50~\cite{he2016deep}, ResNet-101, ResNet-152, VGG16-BN~\cite{simonyan2014very}, and Inception-ResNet-v2 (Inc-Res-v2). On the other hand, the defensively trained models comprise the Ensemble Adversarially Trained Inception-v3 (Inc-v3-ens3), Ensemble Adversarially Trained Inception-v3 with Four Variants (Inc-v3-ens4), and Ensemble Adversarially Trained Inception-ResNet-v2 (Inc-Res-v2-ens). These models integrate multiple adversarial sample versions during the training process, thereby enhancing the models' robustness against adversarial attacks. Furthermore, in our experiments on attack transferability, we selected four models as surrogate models, specifically Inception-v3, Inception-v4, ResNet-152, and Inc-Res-v2.

\subsection{Attack Methods}
To thoroughly assess the effectiveness of the DMS strategy in a white-box attack context, this study initially selected seven attack methods for experimental analysis, including I-FGSM~\cite{kurakin2018adversarial}, PGD~\cite{madry2017towards}, C\&W~\cite{carlini2017towards}, MI-FGSM~\cite{dong2018boosting}, TI-FGSM~\cite{dong2019evading}, SINI-FGSM~\cite{lin2019nesterov}, and SSA~\cite{long2022frequency}.

Furthermore, we assess the performance of DMS on a broader range of adversarial attack methods. Eleven different attack techniques were selected in our experiments, including I-FGSM~\cite{kurakin2018adversarial}, AdvGAN~\cite{xiao2018generating}, DI-FGSM~\cite{xie2019improving}, MI-FGSM~\cite{dong2018boosting}, GE-AdvGAN~\cite{zhu2024geadvgan}, SINI-FGSM~\cite{lin2019nesterov}, NAA~\cite{zhang2022improving}, MIG~\cite{ma2023transferable}, DANAA~\cite{jin2023danaa}, FSPS~\cite{zhu2023improving}, SSA~\cite{long2022frequency}, and GRA~\cite{zhu2023boosting} to thoroughly evaluate the transferability attack performance.

\subsection{Metrics}
The Attack Success Rate (ASR) is selected as the metric to evaluate the performance. It measures the average proportion of mislabeled predicted samples generated by different integerization methods to all generated adversarial samples after the attack. Hence, a higher mean success rate indicates better integerization method to retain the attack performance.

\subsection{Baseline Methods}
We choose the case where the pixel values are still discrete unintegrated values and not saved as images as our baseline (represented by the \textit{Original} legend). In addition, we select the following four competitive methods: upper~\cite{robidoux2008fast}, truncating~\cite{hattne2016modeling}, rounding~\cite{ehsan2015integral} and DMS-AI. The upper method rounds up non-integer pixel values by one. The truncating method removes digits after the decimal point. The rounding method rounds values, while the DMS-AI method performs adversarial integerization only. We compare our DMS method with them under different adversarial approaches and various settings to validate the effectiveness of our method.

\subsection{Parameter Setting}
Following the results from the literature, we have selected the parameters for the fairness experiment performance as follows: 

The thresholds (\textit{Maximum perturbation}) for I-FGSM, MI-FGSM, TI-FGSM, SINI-FGSM, SSA, and PGD are set to 0.3 across all datasets. For C\&W, the threshold is 6 on ImageNet and 1 on CIFAR-100. The decay factor for MI-FGSM is set to 1, while for TI-FGSM it is 0, and for SINI-FGSM, it remains 1. The number of attack steps for all methods is set to 10. The learning rate (lr) for all attack methods is fixed at 0.01, and the number of iterations (T) is set to 100 for C\&W and 50 for I-FGSM, MI-FGSM, TI-FGSM, SINI-FGSM, SSA, and PGD. The diversity\_prob for TI-FGSM is set at 0.5. For SINI-FGSM, the number of random shifts ($m$) is 5. In SSA, the momentum is set to 1.0, $N$ to 20, sigma to 16, and rho to 0.5.

\subsection{Results}

\subsubsection{White-Box Adversarial Attacks}
We firstly compare the performance of different integerization approaches in Table.~\ref{cifar100} and~\ref{imagenet} under seven representative gradient-based attack methods. It is clear that DMS method consistently perform much better than other pixel operation methods for image processing, including truncating, rounding and upper approaches for different attack methods.

\begin{table}[htpb]
\caption{Attack success rate of different methods on CIFAR-100}
\label{cifar100}
\centering
\resizebox{\linewidth}{!}{%
\begin{tabular}{c|c|ccccc}
\hline
Model                         & \multicolumn{1}{c|}{Method} & Truncating & Rounding & Upper   & DMS-AI (ours)       & DMS (ours)          \\ \hline
\multirow{7}{*}{Inception-v3} & I-FGSM                        & 95.58\%    & 97.16\%  & 95.48\% & \textbf{100.00\%} & \textbf{100.00\%} \\
                               & PGD                         & 98.42\%    & 98.21\%  & 98.32\% & \textbf{100.00\%} & \textbf{100.00\%} \\
                               & C\&W                        & 88.64\%    & 24.08\%  & 87.80\% & 99.16\%           & \textbf{99.26\%}  \\
                               & MI-FGSM                     & 95.90\%    & 98.74\%  & 95.58\% & \textbf{100.00\%} & \textbf{100.00\%} \\
                               & SINI-FGSM                    & 87.19\%    & 85.17\%  & 86.94\% & 99.61\%           & \textbf{100.00\%} \\
                               & SSA                         & 88.84\%    & 85.29\%  & 88.71\% & \textbf{100.00\%} & \textbf{100.00\%} \\
                               & TI-FGSM                      & 87.83\%    & 86.43\%  & 86.94\% & \textbf{100.00\%} & \textbf{100.00\%} \\\hline
\multirow{7}{*}{ResNet-50}    & I-FGSM                        & 99.57\%    & 99.89\%  & 99.78\% & \textbf{100.00\%} & \textbf{100.00\%} \\
                               & PGD                         & 97.63\%    & 97.74\%  & 97.63\% & \textbf{100.00\%} & \textbf{100.00\%} \\
                               & C\&W                        & 92.56\%    & 73.17\%  & 92.24\% & \textbf{99.89\%}  & \textbf{99.89\%}  \\
                               & MI-FGSM                     & 99.78\%    & 100.00\% & 99.89\% & \textbf{100.00\%} & \textbf{100.00\%} \\
                               & SINI-FGSM                    & 89.21\%    & 85.62\%  & 87.93\% & 99.74\%           & \textbf{100.00\%} \\
                               & SSA                         & 86.52\%    & 85.10\%  & 85.62\% & 99.87\%           & \textbf{100.00\%} \\
                               & TI-FGSM                      & 87.93\%    & 85.10\%  & 86.39\% & 99.74\%           & \textbf{100.00\%} \\\hline
\multirow{7}{*}{VGG16-BN}     & I-FGSM                        & 99.44\%    & 99.55\%  & 99.44\% & \textbf{100.00\%} & \textbf{100.00\%} \\
                               & PGD                         & 99.55\%    & 99.55\%  & 99.55\% & \textbf{100.00\%} & \textbf{100.00\%} \\
                               & C\&W                        & 94.32\%    & 73.72\%  & 93.65\% & \textbf{100.00\%} & \textbf{100.00\%} \\
                               & MI-FGSM                     & 99.33\%    & 99.44\%  & 99.33\% & \textbf{100.00\%} & \textbf{100.00\%} \\ 
                               & SINI-FGSM                    & 88.01\%    & 84.62\%  & 86.88\% & 99.85\%           & \textbf{100.00\%} \\
                               & SSA                         & 88.01\%    & 86.88\%  & 86.17\% & 98.87\%  & \textbf{99.43\%} \\
                               & TI-FGSM                      & 89.42\%    & 88.57\%  & 87.30\% & \textbf{98.73\%}  & \textbf{98.73\%} \\\hline
\end{tabular}
}
\end{table}

In particular, we evaluate the attack success rates of seven different attack methods (I-FGSM, TI-FGSM, SINI-FGSM, MI-FGSM, PGD, C\&W, and SSA) on CIFAR-100 and ImageNet datasets. As shown in Table.~\ref{cifar100}, we can see that the performance fluctuations for the traditional pixel storage operations are large in terms of the attacking performance. None of the operations can guarantee the obtained adversarial samples can retain as much attack performance as they present during the training step. However, for the DMS method, the experiment demonstrates the best performance in terms of attack success rate, surpassing all other traditional approaches (Rounding, Truncating, and Upper). Notably, the DMS method further boosts the enhanced performance of DMS-AI to 100.00\% across almost all tasks. In the meantime, the performance gap between DMS-AI and DMS is surprisingly small for the CIFAR-100 dataset, which may be related to the relatively simple and small nature of the dataset.

\begin{table}[htpb]
\caption{Attack success rate of different methods on ImageNet}
\label{imagenet}
\centering
\resizebox{\linewidth}{!}{%
\begin{tabular}{c|c|ccccc}
\hline
Model                          & Method         & Truncating & Rounding & Upper   & DMS-AI (our)            & DMS (our)               \\ \hline
\multirow{7}{*}{Inception-v3} & I-FGSM           & 90.87\%    & 86.19\%  & 89.99\% & \textbf{100.00\%} & \textbf{100.00\%} \\
                               & PGD            & 96.07\%    & 96.07\%  & 95.82\% & \textbf{100.00\%} & \textbf{100.00\%} \\
                               & C\&W           & 82.00\%    & 24.59\%  & 81.37\% & 96.32\%           & \textbf{97.59\%}  \\
                               & MI-FGSM        & 89.86\%    & 86.82\%  & 89.23\% & \textbf{100.00\%} & \textbf{100.00\%} \\ 
                               & SINI-FGSM       & 90.53\%    & 91.58\%  & 90.22\% & \textbf{100.00\%} & \textbf{100.00\%} \\
                               & SSA            & 88.32\%    & 85.06\%  & 88.22\% & \textbf{100.00\%} & \textbf{100.00\%} \\
                               & TI-FGSM         & 90.95\%    & 91.79\%  & 91.06\% & \textbf{100.00\%} & \textbf{100.00\%} \\\hline
\multirow{7}{*}{ResNet-50}    & I-FGSM           & 89.60\%    & 87.80\%  & 88.58\% & \textbf{99.87\%}  & \textbf{99.87\%}  \\
                               & PGD            & 96.79\%    & 96.66\%  & 96.41\% & 99.61\%           & \textbf{100.00\%} \\
                               & C\&W           & 78.95\%    & 27.21\%  & 79.20\% & 96.53\%           & \textbf{98.07\%}  \\
                               & MI-FGSM        & 89.22\%    & 88.70\%  & 88.06\% & 99.87\%           & \textbf{100.00\%} \\
                               & SINI-FGSM       & 98.59\%    & 98.81\%  & 98.59\% & \textbf{100.00\%} & \textbf{100.00\%} \\
                               & SSA            & 97.95\%    & 98.16\%  & 97.41\% & \textbf{100.00\%} & \textbf{100.00\%} \\
                               & TI-FGSM         & 97.95\%    & 98.06\%  & 97.19\% & \textbf{100.00\%} & \textbf{100.00\%} \\\hline
\multirow{7}{*}{VGG16-BN}     & I-FGSM           & 89.00\%    & 90.83\%  & 87.73\% & \textbf{99.72\%}  & \textbf{99.72\%}  \\
                               & PGD            & 95.77\%    & 95.49\%  & 95.35\% & 99.29\%           & \textbf{99.86\%}  \\
                               & C\&W           & 75.32\%    & 33.71\%  & 74.05\% & 92.67\%           & \textbf{95.49\%}  \\
                               & MI-FGSM        & 87.59\%    & 89.99\%  & 86.46\% & 99.29\%           & \textbf{99.44\%} \\
                               & SINI-FGSM       & 98.55\%    & 98.77\%  & 98.55\% & \textbf{100.00\%} & \textbf{100.00\%} \\
                               & SSA            & 98.99\%    & 98.33\%  & 98.88\% & \textbf{100.00\%} & \textbf{100.00\%} \\
                               & TI-FGSM         & 98.88\%    & 98.32\%  & 98.66\% & \textbf{100.00\%} & \textbf{100.00\%} \\\hline
\end{tabular}
}
\end{table}

\begin{figure*}[htbp]
  \centering
  % \vspace{-3cm}
  % \includegraphics[scale=0.11]{images/diff_lr.pdf}
  \includegraphics[width=\textwidth]{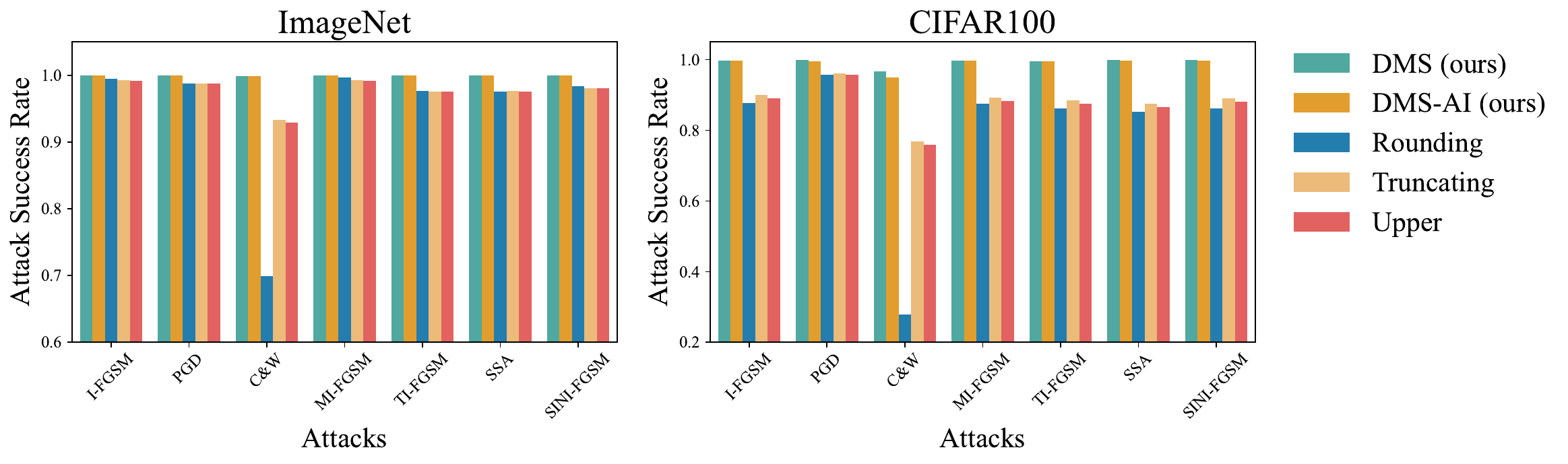}
  \caption{Mean success rate of different integerization methods on ImageNet}
  \label{fig:ASR}
\end{figure*}

Thus, for the ImageNet dataset, as illustrated in Figure.~\ref{fig:ASR}, the performance gap between DMS and other methods becomes more pronounced. As shown in Table.~\ref{imagenet}, under the C\&W method, the maximum performance gap between DMS and other methods can be as much as 73\%, i.e., between DMS and Rounding for Inception-v3. Additionally, the average maximum performance gap for the C\&W method is 36.28\%. Considering all scenarios, DMS exhibits an average improvement of 9.71\% compared to the other methods. Given ImageNet dataset is a larger and more diverse dataset than CIFAR-100, we can see that the proposed DMS method has consistently demonstrated the most stable and highest performance throughout all the benchmarking results.

\subsubsection{Transferable Adversarial Attacks}

\paragraph{Inception-v3 and Inception-v4 as the Surrogate Model} 

In this section, we evaluated the DMS method against three other techniques—Truncating, Rounding, and Upper—using Inception-v3 and Inception-v4 as surrogate models. Our results, summarized in Table 1, demonstrate that DMS surpasses its counterparts in 854 out of 864 attack scenarios, yielding an impressive efficacy rate of 98.84\%. Notably, DMS consistently improves the ASR, averaging a 1.03\% enhancement across all scenarios.

Particularly noteworthy is the performance of DMS in attacks involving Generative Structures, such as AdvGAN and GE-AdvGAN, when using Inception-v3 as the surrogate model. Here, DMS exhibits a significant ASR increase of 2.73\% and 1.83\%, respectively, over the other methods. This superior performance is likely attributable to the less restrictive nature of generative structures regarding remainder terms during attacks. Unlike the other techniques, DMS operates with continuous output values, only limited by computational precision, allowing for a wider range of remainder options. This factor is crucial in augmenting the method's effectiveness in generative-style attacks.

Under the Inception-v4 model, DMS maintains its high efficacy in similar attack scenarios, recording ASR improvements of 2.18\% and 1.94\% for AdvGAN and GE-AdvGAN, respectively. Additionally, there appears to be a positive correlation between the strength of transferability and the improvement margin provided by DMS. For instance, in methods like DANAA, NAA, GRA, SSA, and FSPS, DMS averages increases of 1.39\%, 0.92\%, 0.98\%, 0.80\%, and 0.76\%, in that order. However, this improvement margin is slightly less pronounced in the case of I-FGSM, where it stands at 0.54\%.

\begin{table*}[htpb]
\centering
\caption{Comparative Analysis of Attack Success Rates using Inception-v3 and Inception-v4 as Surrogate Model. This table presents a comprehensive evaluation of various attack methods across multiple models.}
\label{tab:Inc-v3-v4}
\resizebox{\textwidth}{!}{%
\begin{tabular}{@{}cc|ccccccccc|ccccccccc@{}}
\toprule
\multicolumn{2}{c|}{Surrogate Model}                         & \multicolumn{9}{c|}{Inc-v3}                                                                                                                                                                                                                                                                   & \multicolumn{9}{c}{Inc-v4}                                                                                                                                                                                                                                                                    \\ \midrule
\multicolumn{1}{c|}{Attack}                     & Method     & Inc-v4           & Inc-Res-v2       & Res-50           & Res-101          & Res-152          & \begin{tabular}[c]{@{}c@{}}Inc-v3\\ -ens3\end{tabular} & \begin{tabular}[c]{@{}c@{}}Inc-v3\\ -ens4\end{tabular} & \begin{tabular}[c]{@{}c@{}}Inc-Res\\ -v2-ens\end{tabular} & Average          & Inc-v3           & Inc-Res-v2       & Res-50           & Res-101          & Res-152          & \begin{tabular}[c]{@{}c@{}}Inc-v3\\ -ens3\end{tabular} & \begin{tabular}[c]{@{}c@{}}Inc-v3\\ -ens4\end{tabular} & \begin{tabular}[c]{@{}c@{}}Inc-Res\\ -v2-ens\end{tabular} & Average          \\ \midrule
\multicolumn{1}{c|}{\multirow{4}{*}{I-FGSM}}       & DMS        & \textbf{28.80\%} & \textbf{20.80\%} & 24.00\%          & \textbf{22.20\%} & \textbf{19.10\%} & \textbf{12.70\%}                                       & 12.80\%                                                & \textbf{5.10\%}                                           & \textbf{18.19\%} & \textbf{37.50\%} & \textbf{19.20\%} & \textbf{24.10\%} & \textbf{20.60\%} & 20.00\%          & 11.30\%                                                & \textbf{11.40\%}                                       & \textbf{5.20\%}                                           & \textbf{18.66\%} \\
\multicolumn{1}{c|}{}                           & Truncating & 27.90\%          & 19.70\%          & \textbf{24.10\%} & 21.90\%          & 18.70\%          & 12.50\%                                                & 13.10\%                                                & 4.70\%                                                    & 17.83\%          & 36.10\%          & 18.90\%          & 23.90\%          & 20.00\%          & \textbf{20.10\%} & 11.30\%                                                & 11.00\%                                                & 5.10\%                                                    & 18.30\%          \\
\multicolumn{1}{c|}{}                           & Rounding   & 26.90\%          & 19.10\%          & 22.60\%          & 21.00\%          & 18.40\%          & 12.50\%                                                & \textbf{13.30\%}                                       & 5.00\%                                                    & 17.35\%          & 34.90\%          & 18.10\%          & 23.30\%          & 19.10\%          & 19.00\%          & 11.20\%                                                & \textbf{11.40\%}                                       & \textbf{5.20\%}                                           & 17.78\%          \\
\multicolumn{1}{c|}{}                           & Upper      & 28.00\%          & 19.40\%          & 23.90\%          & 21.80\%          & 18.70\%          & 12.70\%                                                & 13.10\%                                                & 4.60\%                                                    & 17.78\%          & 36.20\%          & 18.90\%          & 23.80\%          & 20.10\%          & 19.90\%          & \textbf{11.40\%}                                       & 11.00\%                                                & 5.10\%                                                    & 18.30\%          \\ \midrule
\multicolumn{1}{c|}{\multirow{4}{*}{AdvGAN}}    & DMS        & \textbf{51.30\%} & \textbf{31.20\%} & \textbf{69.70\%} & \textbf{64.10\%} & \textbf{49.70\%} & \textbf{44.30\%}                                       & \textbf{41.00\%}                                       & \textbf{11.20\%}                                          & \textbf{45.31\%} & \textbf{66.50\%} & \textbf{14.70\%} & \textbf{60.60\%} & \textbf{61.60\%} & \textbf{46.50\%} & \textbf{18.90\%}                                       & \textbf{33.90\%}                                       & \textbf{15.40\%}                                          & \textbf{39.76\%} \\
\multicolumn{1}{c|}{}                           & Truncating & 48.50\%          & 29.50\%          & 66.40\%          & 61.20\%          & 45.60\%          & 41.40\%                                                & 37.90\%                                                & 10.40\%                                                   & 42.61\%          & 63.80\%          & 13.70\%          & 57.60\%          & 59.00\%          & 43.30\%          & 17.50\%                                                & 32.00\%                                                & 14.50\%                                                   & 37.68\%          \\
\multicolumn{1}{c|}{}                           & Rounding   & 48.60\%          & 29.40\%          & 66.40\%          & 61.00\%          & 45.50\%          & 41.90\%                                                & 37.40\%                                                & 10.30\%                                                   & 42.56\%          & 63.60\%          & 13.70\%          & 57.60\%          & 58.90\%          & 43.40\%          & 17.10\%                                                & 31.70\%                                                & 14.40\%                                                   & 37.55\%          \\
\multicolumn{1}{c|}{}                           & Upper      & 48.40\%          & 29.30\%          & 66.50\%          & 61.10\%          & 45.50\%          & 41.70\%                                                & 37.70\%                                                & 10.30\%                                                   & 42.56\%          & 63.70\%          & 13.70\%          & 57.20\%          & 58.90\%          & 43.10\%          & 17.20\%                                                & 31.90\%                                                & 14.50\%                                                   & 37.53\%          \\ \midrule
\multicolumn{1}{c|}{\multirow{4}{*}{DI-FGSM}}   & DMS        & \textbf{46.40\%} & \textbf{39.40\%} & \textbf{38.30\%} & \textbf{33.80\%} & \textbf{31.50\%} & 17.90\%                                                & 17.80\%                                                & \textbf{8.00\%}                                           & \textbf{29.14\%} & \textbf{57.00\%} & \textbf{39.50\%} & \textbf{37.80\%} & \textbf{33.20\%} & \textbf{33.00\%} & 14.80\%                                                & 16.40\%                                                & 7.90\%                                                    & \textbf{29.95\%} \\
\multicolumn{1}{c|}{}                           & Truncating & 44.70\%          & 37.80\%          & 37.40\%          & 33.40\%          & 30.30\%          & 17.80\%                                                & 18.20\%                                                & \textbf{8.00\%}                                           & 28.45\%          & 55.80\%          & 38.60\%          & 37.10\%          & 33.00\%          & 32.70\%          & 14.50\%                                                & 15.90\%                                                & 8.00\%                                                    & 29.45\%          \\
\multicolumn{1}{c|}{}                           & Rounding   & 44.20\%          & 36.20\%          & 36.30\%          & 31.90\%          & 29.30\%          & \textbf{18.00\%}                                       & \textbf{18.30\%}                                       & 7.80\%                                                    & 27.75\%          & 55.40\%          & 37.30\%          & 35.80\%          & 32.90\%          & 32.10\%          & \textbf{15.30\%}                                       & \textbf{16.50\%}                                       & \textbf{8.20\%}                                           & 29.19\%          \\
\multicolumn{1}{c|}{}                           & Upper      & 44.70\%          & 37.80\%          & 37.10\%          & 33.60\%          & 30.10\%          & 17.80\%                                                & \textbf{18.30\%}                                       & 7.70\%                                                    & 28.39\%          & 55.70\%          & 38.50\%          & 37.30\%          & 33.20\%          & 32.80\%          & 14.60\%                                                & 16.00\%                                                & 8.10\%                                                    & 29.53\%          \\ \midrule
\multicolumn{1}{c|}{\multirow{4}{*}{MI-FGSM}}   & DMS        & \textbf{50.40\%} & \textbf{47.60\%} & \textbf{47.00\%} & \textbf{42.40\%} & \textbf{42.50\%} & \textbf{22.70\%}                                       & 23.10\%                                                & \textbf{10.90\%}                                          & \textbf{35.83\%} & \textbf{62.40\%} & \textbf{46.60\%} & \textbf{45.90\%} & \textbf{43.70\%} & \textbf{42.50\%} & \textbf{19.20\%}                                       & \textbf{20.30\%}                                       & \textbf{11.20\%}                                          & \textbf{36.48\%} \\
\multicolumn{1}{c|}{}                           & Truncating & 50.20\%          & 46.50\%          & \textbf{47.00\%} & 41.30\%          & 41.50\%          & 22.50\%                                                & 22.50\%                                                & 10.50\%                                                   & 35.25\%          & 61.20\%          & 46.10\%          & 45.60\%          & 42.20\%          & 42.10\%          & 19.10\%                                                & 19.90\%                                                & \textbf{11.20\%}                                          & 35.93\%          \\
\multicolumn{1}{c|}{}                           & Rounding   & \textbf{50.40\%} & 47.00\%          & 46.90\%          & 41.50\%          & 41.60\%          & 22.20\%                                                & \textbf{23.30\%}                                       & 10.60\%                                                   & 35.44\%          & 61.70\%          & 46.80\%          & 46.00\%          & 42.70\%          & 42.20\%          & 19.00\%                                                & 19.70\%                                                & 11.10\%                                                   & 36.15\%          \\
\multicolumn{1}{c|}{}                           & Upper      & 50.00\%          & 45.90\%          & \textbf{47.00\%} & 41.40\%          & 41.30\%          & 22.20\%                                                & 22.70\%                                                & 10.60\%                                                   & 35.14\%          & 61.10\%          & 45.90\%          & 45.30\%          & 42.10\%          & 41.80\%          & 19.00\%                                                & 20.00\%                                                & 11.20\%                                                   & 35.80\%          \\ \midrule
\multicolumn{1}{c|}{\multirow{4}{*}{GE-AdvGAN}} & DMS        & \textbf{70.10\%} & \textbf{48.80\%} & \textbf{79.70\%} & \textbf{82.90\%} & \textbf{75.20\%} & \textbf{53.20\%}                                       & \textbf{62.80\%}                                       & \textbf{33.70\%}                                          & \textbf{63.30\%} & \textbf{79.10\%} & \textbf{34.50\%} & \textbf{82.50\%} & \textbf{87.90\%} & \textbf{77.40\%} & \textbf{23.80\%}                                       & \textbf{47.90\%}                                       & \textbf{24.30\%}                                          & \textbf{57.18\%} \\
\multicolumn{1}{c|}{}                           & Truncating & 68.80\%          & 47.50\%          & 78.20\%          & 82.10\%          & 73.40\%          & 50.70\%                                                & 59.80\%                                                & 31.80\%                                                   & 61.54\%          & 77.20\%          & 32.10\%          & 80.80\%          & 86.20\%          & 75.70\%          & 22.70\%                                                & 45.60\%                                                & 22.60\%                                                   & 55.36\%          \\
\multicolumn{1}{c|}{}                           & Rounding   & 68.80\%          & 47.30\%          & 78.10\%          & 82.10\%          & 73.20\%          & 50.10\%                                                & 59.60\%                                                & 31.70\%                                                   & 61.36\%          & 77.20\%          & 31.80\%          & 80.60\%          & 86.10\%          & 75.70\%          & 22.10\%                                                & 44.80\%                                                & 22.60\%                                                   & 55.11\%          \\
\multicolumn{1}{c|}{}                           & Upper      & 68.80\%          & 47.50\%          & 78.20\%          & 81.80\%          & 73.40\%          & 50.80\%                                                & 59.70\%                                                & 32.00\%                                                   & 61.53\%          & 77.10\%          & 32.00\%          & 80.60\%          & 86.00\%          & 75.80\%          & 22.60\%                                                & 45.10\%                                                & 22.60\%                                                   & 55.23\%          \\ \midrule
\multicolumn{1}{c|}{\multirow{4}{*}{SINI-FGSM}} & DMS        & \textbf{77.80\%} & \textbf{76.00\%} & \textbf{75.00\%} & \textbf{69.80\%} & \textbf{68.40\%} & \textbf{40.90\%}                                       & \textbf{37.70\%}                                       & \textbf{23.90\%}                                          & \textbf{58.69\%} & \textbf{87.70\%} & \textbf{79.80\%} & \textbf{77.20\%} & \textbf{73.90\%} & \textbf{74.40\%} & \textbf{49.30\%}                                       & \textbf{44.60\%}                                       & \textbf{30.10\%}                                          & \textbf{64.63\%} \\
\multicolumn{1}{c|}{}                           & Truncating & 76.60\%          & 75.50\%          & 73.20\%          & 69.00\%          & 66.90\%          & 39.60\%                                                & 36.70\%                                                & 23.00\%                                                   & 57.56\%          & 87.40\%          & 78.90\%          & 76.90\%          & 73.30\%          & 73.10\%          & 48.70\%                                                & 43.60\%                                                & 29.60\%                                                   & 63.94\%          \\
\multicolumn{1}{c|}{}                           & Rounding   & 76.80\%          & 75.70\%          & 73.90\%          & 69.50\%          & 67.50\%          & 39.90\%                                                & 37.00\%                                                & 23.10\%                                                   & 57.93\%          & 87.50\%          & 79.60\%          & 77.00\%          & 73.70\%          & 73.80\%          & 48.30\%                                                & 44.20\%                                                & 29.40\%                                                   & 64.19\%          \\
\multicolumn{1}{c|}{}                           & Upper      & 76.40\%          & 75.50\%          & 73.00\%          & 69.00\%          & 66.80\%          & 39.60\%                                                & 36.60\%                                                & 23.00\%                                                   & 57.49\%          & 87.20\%          & 78.80\%          & 76.90\%          & 72.90\%          & 72.80\%          & 48.60\%                                                & 43.60\%                                                & 29.50\%                                                   & 63.79\%          \\ \midrule
\multicolumn{1}{c|}{\multirow{4}{*}{NAA}}       & DMS        & \textbf{88.50\%} & \textbf{87.20\%} & \textbf{83.50\%} & \textbf{83.70\%} & \textbf{82.50\%} & \textbf{59.70\%}                                       & \textbf{57.50\%}                                       & \textbf{36.50\%}                                          & \textbf{72.39\%} & \textbf{88.40\%} & \textbf{84.10\%} & \textbf{82.70\%} & \textbf{83.00\%} & \textbf{80.90\%} & \textbf{62.50\%}                                       & \textbf{58.20\%}                                       & \textbf{42.60\%}                                          & \textbf{72.80\%} \\
\multicolumn{1}{c|}{}                           & Truncating & 87.60\%          & 86.20\%          & 82.80\%          & 83.10\%          & 81.50\%          & 57.60\%                                                & 55.40\%                                                & 34.70\%                                                   & 71.11\%          & 88.20\%          & 83.40\%          & 82.50\%          & 82.40\%          & 80.20\%          & 60.00\%                                                & 56.70\%                                                & 41.20\%                                                   & 71.83\%          \\
\multicolumn{1}{c|}{}                           & Rounding   & 87.80\%          & 86.50\%          & 83.10\%          & 83.40\%          & 81.70\%          & 58.40\%                                                & 55.70\%                                                & 35.30\%                                                   & 71.49\%          & 88.10\%          & 83.50\%          & 82.30\%          & 82.70\%          & 80.50\%          & 60.80\%                                                & 56.80\%                                                & 41.50\%                                                   & 72.03\%          \\
\multicolumn{1}{c|}{}                           & Upper      & 87.60\%          & 86.10\%          & 82.90\%          & 83.00\%          & 81.30\%          & 57.60\%                                                & 55.50\%                                                & 34.80\%                                                   & 71.10\%          & 88.20\%          & 83.40\%          & 82.40\%          & 82.30\%          & 80.20\%          & 60.20\%                                                & 56.50\%                                                & 41.10\%                                                   & 71.79\%          \\ \midrule
\multicolumn{1}{c|}{\multirow{4}{*}{MIG}}       & DMS        & \textbf{71.50\%} & \textbf{70.20\%} & \textbf{70.20\%} & \textbf{65.90\%} & \textbf{64.70\%} & \textbf{40.80\%}                                       & \textbf{40.70\%}                                       & \textbf{21.80\%}                                          & \textbf{55.73\%} & \textbf{86.40\%} & 79.80\%          & 77.40\%          & \textbf{75.20\%} & \textbf{75.40\%} & \textbf{52.90\%}                                       & \textbf{49.60\%}                                       & \textbf{34.50\%}                                          & \textbf{66.40\%} \\
\multicolumn{1}{c|}{}                           & Truncating & 70.80\%          & 68.50\%          & 69.80\%          & 64.70\%          & 63.90\%          & 40.20\%                                                & 39.60\%                                                & 21.30\%                                                   & 54.85\%          & 86.30\%          & 79.60\%          & 77.30\%          & 74.80\%          & 74.50\%          & 51.40\%                                                & 48.90\%                                                & 34.40\%                                                   & 65.90\%          \\
\multicolumn{1}{c|}{}                           & Rounding   & 71.20\%          & 69.10\%          & 69.90\%          & 65.30\%          & 64.40\%          & 40.20\%                                                & 40.10\%                                                & 21.70\%                                                   & 55.24\%          & 86.30\%          & \textbf{80.00\%} & \textbf{77.60\%} & 75.20\%          & 75.00\%          & 52.10\%                                                & 49.20\%                                                & 34.20\%                                                   & 66.20\%          \\
\multicolumn{1}{c|}{}                           & Upper      & 70.30\%          & 68.60\%          & 69.80\%          & 64.40\%          & 63.60\%          & 40.30\%                                                & 39.80\%                                                & 21.30\%                                                   & 54.76\%          & 86.20\%          & 79.60\%          & 77.20\%          & 74.80\%          & 74.20\%          & 51.50\%                                                & 48.90\%                                                & 34.40\%                                                   & 65.85\%          \\ \midrule
\multicolumn{1}{c|}{\multirow{4}{*}{DANAA}}     & DMS        & \textbf{92.60\%} & \textbf{91.40\%} & \textbf{90.30\%} & \textbf{89.70\%} & \textbf{88.80\%} & \textbf{69.10\%}                                       & \textbf{66.60\%}                                       & \textbf{49.40\%}                                          & \textbf{79.74\%} & \textbf{93.70\%} & \textbf{90.90\%} & \textbf{89.30\%} & \textbf{88.80\%} & \textbf{87.90\%} & \textbf{74.00\%}                                       & \textbf{69.10\%}                                       & \textbf{52.10\%}                                          & \textbf{80.73\%} \\
\multicolumn{1}{c|}{}                           & Truncating & 92.10\%          & 91.20\%          & 90.00\%          & 89.10\%          & 88.10\%          & 67.40\%                                                & 64.40\%                                                & 47.00\%                                                   & 78.66\%          & 92.90\%          & 89.40\%          & 88.50\%          & 88.10\%          & 87.40\%          & 72.60\%                                                & 67.80\%                                                & 48.90\%                                                   & 79.45\%          \\
\multicolumn{1}{c|}{}                           & Rounding   & 92.00\%          & 91.10\%          & 90.00\%          & 89.10\%          & 88.00\%          & 67.40\%                                                & 64.10\%                                                & 46.80\%                                                   & 78.56\%          & 92.80\%          & 89.40\%          & 88.40\%          & 88.10\%          & 87.40\%          & 72.10\%                                                & 67.30\%                                                & 48.40\%                                                   & 79.24\%          \\
\multicolumn{1}{c|}{}                           & Upper      & 92.10\%          & 91.00\%          & 90.00\%          & 89.20\%          & 88.00\%          & 67.10\%                                                & 64.30\%                                                & 46.90\%                                                   & 78.58\%          & 92.60\%          & 89.40\%          & 88.50\%          & 88.40\%          & 87.20\%          & 72.50\%                                                & 67.70\%                                                & 48.30\%                                                   & 79.33\%          \\ \midrule
\multicolumn{1}{c|}{\multirow{4}{*}{FSPS}}      & DMS        & \textbf{92.00\%} & \textbf{91.20\%} & \textbf{88.40\%} & \textbf{88.00\%} & \textbf{88.60\%} & \textbf{84.40\%}                                       & \textbf{81.60\%}                                       & \textbf{73.50\%}                                          & \textbf{85.96\%} & \textbf{92.70\%} & \textbf{90.20\%} & \textbf{88.10\%} & \textbf{86.40\%} & \textbf{87.70\%} & \textbf{81.40\%}                                       & \textbf{79.20\%}                                       & \textbf{73.50\%}                                          & \textbf{84.90\%} \\
\multicolumn{1}{c|}{}                           & Truncating & 91.80\%          & 90.90\%          & 87.90\%          & 86.80\%          & 87.80\%          & 82.90\%                                                & 80.70\%                                                & 71.60\%                                                   & 85.05\%          & 92.30\%          & 89.90\%          & 87.50\%          & 85.30\%          & 86.90\%          & 80.80\%                                                & 78.30\%                                                & 71.80\%                                                   & 84.10\%          \\
\multicolumn{1}{c|}{}                           & Rounding   & 91.80\%          & 90.90\%          & 88.30\%          & 87.50\%          & 88.10\%          & 83.30\%                                                & 81.00\%                                                & 72.50\%                                                   & 85.43\%          & 92.40\%          & 90.10\%          & 87.70\%          & 85.60\%          & 87.00\%          & 81.00\%                                                & 78.40\%                                                & 71.90\%                                                   & 84.26\%          \\
\multicolumn{1}{c|}{}                           & Upper      & 91.80\%          & 90.80\%          & 87.80\%          & 86.70\%          & 87.80\%          & 82.80\%                                                & 80.30\%                                                & 72.00\%                                                   & 85.00\%          & 92.30\%          & 89.90\%          & 87.60\%          & 85.10\%          & 86.90\%          & 80.80\%                                                & 78.40\%                                                & 71.50\%                                                   & 84.06\%          \\ \midrule
\multicolumn{1}{c|}{\multirow{4}{*}{SSA}}       & DMS        & \textbf{89.50\%} & \textbf{88.00\%} & \textbf{83.30\%} & \textbf{81.90\%} & \textbf{81.90\%} & \textbf{75.90\%}                                       & \textbf{73.90\%}                                       & \textbf{62.50\%}                                          & \textbf{79.61\%} & \textbf{91.40\%} & \textbf{87.10\%} & \textbf{84.50\%} & \textbf{82.40\%} & \textbf{82.20\%} & \textbf{75.30\%}                                       & \textbf{74.10\%}                                       & \textbf{65.20\%}                                          & \textbf{80.28\%} \\
\multicolumn{1}{c|}{}                           & Truncating & 88.90\%          & 87.10\%          & 82.50\%          & 81.20\%          & 81.10\%          & 75.10\%                                                & 72.70\%                                                & 61.40\%                                                   & 78.75\%          & 91.20\%          & 86.50\%          & 83.40\%          & 81.20\%          & 81.60\%          & 73.80\%                                                & 73.20\%                                                & 63.80\%                                                   & 79.34\%          \\
\multicolumn{1}{c|}{}                           & Rounding   & 89.20\%          & 87.40\%          & 82.70\%          & 81.50\%          & 81.40\%          & 75.40\%                                                & 73.60\%                                                & 61.50\%                                                   & 79.09\%          & 91.30\%          & 86.80\%          & 83.70\%          & 81.50\%          & 81.90\%          & 74.60\%                                                & 73.70\%                                                & 64.40\%                                                   & 79.74\%          \\
\multicolumn{1}{c|}{}                           & Upper      & 88.90\%          & 87.00\%          & 82.50\%          & 81.10\%          & 81.10\%          & 75.10\%                                                & 72.90\%                                                & 61.30\%                                                   & 78.74\%          & 91.30\%          & 86.40\%          & 83.30\%          & 81.20\%          & 81.50\%          & 74.00\%                                                & 73.30\%                                                & 63.70\%                                                   & 79.34\%          \\ \midrule
\multicolumn{1}{c|}{\multirow{4}{*}{GRA}}      & DMS        & \textbf{88.10\%}               & \textbf{86.90\%}                   & \textbf{83.10\%}                  & \textbf{81.90\%}                   & \textbf{82.00\%}                   & \textbf{67.30\%}                    & \textbf{66.70\%}                    & \textbf{47.60\%}                       & \textbf{75.45\%} & \textbf{90.60\%}               & \textbf{85.90\%}                   & \textbf{81.90\%}                  & \textbf{80.80\%}                   & \textbf{81.50\%}                   & \textbf{69.20\%}                    & \textbf{66.80\%}                    & \textbf{54.20\%}                       & \textbf{76.36\%} \\
\multicolumn{1}{c|}{}      & Truncating & 87.30\%                        & 86.30\%                            & 82.40\%                           & 81.00\%                            & 81.60\%                            & 66.70\%                             & 66.20\%                             & 46.60\%                                & 74.76\%          & 89.80\%                        & 84.80\%                            & 81.30\%                           & 80.10\%                            & 80.50\%                            & 67.80\%                             & 65.60\%                             & 52.60\%                                & 75.31\%          \\
\multicolumn{1}{c|}{}      & Rounding   & 87.80\%                        & 86.50\%                            & 82.80\%                           & 81.40\%                            & 82.00\%                            & 66.80\%                             & 65.90\%                             & 46.80\%                                & 75.00\%          & 89.80\%                        & 85.30\%                            & 81.50\%                           & 80.20\%                            & 81.00\%                            & 68.40\%                             & 65.90\%                             & 52.70\%                                & 75.60\%          \\
\multicolumn{1}{c|}{} & Upper      & 87.20\%                        & 86.20\%                            & 82.20\%                           & 81.00\%                            & 81.60\%                            & 66.60\%                             & 66.00\%                             & 46.30\%                                & 74.64\%          & 89.80\%                        & 84.80\%                            & 81.30\%                           & 80.00\%                            & 80.50\%                            & 67.90\%                             & 65.50\%                             & 52.60\%                                & 75.30\%          \\ \bottomrule
\end{tabular}%
}
\end{table*}

\paragraph{Inc-Res-v2 and ResNet-152 as the Surrogate Model}
In this section, we focused on evaluating the transferability performance of the Do More Steps (DMS) method in comparison to three other techniques: Truncating, Rounding, and Upper. For this purpose, we employed Inc-Res-v2 and ResNet-152 as surrogate models. As indicated in Table 2, DMS outperformed the other methods in 856 out of 864 attack scenarios, achieving an optimal effectiveness rate of 99.07\%. Furthermore, DMS successfully enhanced the ASR across all methods, averaging a 1.09\% improvement in all attack scenarios.

When utilizing Inc-Res-v2 as the surrogate model, DMS exhibited a consistent performance advantage, especially in attacks involving Generative Structures such as AdvGAN and GE-AdvGAN, similar to its performance with other surrogate models. However, the margin of improvement in these scenarios was relatively more stable compared to other models. Apart from the highest gains in GE-AdvGAN and I-FGSM, the transferability ASR improvements for other attack methods were approximately 1\%.

Conversely, under the ResNet-152 model, the performance of the DMS method showed considerable sharpness. The improvement in the most effective scenario, AdvGAN, was 10.4 times higher than in the least effective one, I-FGSM. Similarly, the second-highest improvement in GE-AdvGAN was 3.43 times greater than that in the second-least effective method, MI-FGSM.

In summary, our research indicates that the DMS method generally enhances the performance of adversarial attack algorithms, particularly in terms of transferability. This finding underscores the potential of DMS in improving the robustness and efficacy of adversarial attacks in various AI applications.

% \usepackage{booktabs}
% \usepackage{multirow}
% \usepackage{graphicx}
\begin{table*}[htpb]
\centering
\caption{Comparative Analysis of Attack Success Rates using Inc-Res-v2 and ResNet-152 as Surrogate Model. This table presents a comprehensive evaluation of various attack methods across multiple models.}
\label{tab:Inc-Res-152}
\resizebox{\textwidth}{!}{%
\begin{tabular}{@{}cc|ccccccccc|ccccccccc@{}}
\toprule
\multicolumn{2}{c|}{Surrogate Model}                         & \multicolumn{9}{c|}{Inc-Res-v2}                                                                                                                                                                                                                                                               & \multicolumn{9}{c}{ResNet-152}                                                                                                                                                                                                                                                                \\ \midrule
\multicolumn{1}{c|}{Attack}                     & Method     & Inc-v3           & Inc-v4           & Res-50           & Res-101          & Res-152          & \begin{tabular}[c]{@{}c@{}}Inc-v3\\ -ens3\end{tabular} & \begin{tabular}[c]{@{}c@{}}Inc-v3\\ -ens4\end{tabular} & \begin{tabular}[c]{@{}c@{}}Inc-Res\\ -v2-ens\end{tabular} & Average          & Inc-v3           & Inc-v4           & Inc-Res-v2       & Res-50           & Res-101          & \begin{tabular}[c]{@{}c@{}}Inc-v3\\ -ens3\end{tabular} & \begin{tabular}[c]{@{}c@{}}Inc-v3\\ -ens4\end{tabular} & \begin{tabular}[c]{@{}c@{}}Inc-Res\\ -v2-ens\end{tabular} & Average          \\ \midrule
\multicolumn{1}{c|}{\multirow{4}{*}{I-FGSM}}       & DMS        & \textbf{33.70\%} & \textbf{29.10\%} & \textbf{24.50\%} & \textbf{21.30\%} & \textbf{20.40\%} & 10.10\%                                                & 11.60\%                                                & 6.90\%                                                    & \textbf{19.70\%} & \textbf{27.30\%} & \textbf{21.40\%} & \textbf{12.30\%} & 29.60\%          & 26.20\%          & 11.30\%                                                & \textbf{12.20\%}                                       & \textbf{4.90\%}                                           & \textbf{18.15\%} \\
\multicolumn{1}{c|}{}                           & Truncating & 33.70\%          & 27.70\%          & 24.30\%          & 21.00\%          & 20.30\%          & \textbf{10.50\%}                                       & 11.50\%                                                & 7.10\%                                                    & 19.51\%          & 27.40\%          & 21.10\%          & 11.90\%          & 29.60\%          & 26.30\%          & 11.30\%                                                & 11.70\%                                                & \textbf{4.90\%}                                           & 18.03\%          \\
\multicolumn{1}{c|}{}                           & Rounding   & 32.50\%          & 26.80\%          & 24.20\%          & 20.30\%          & 20.20\%          & 10.00\%                                                & \textbf{12.00\%}                                       & \textbf{7.30\%}                                           & 19.16\%          & 26.20\%          & 20.90\%          & 10.90\%          & 28.70\%          & 25.90\%          & 11.20\%                                                & 12.00\%                                                & 4.80\%                                                    & 17.58\%          \\
\multicolumn{1}{c|}{}                           & Upper      & 33.60\%          & 27.80\%          & 24.30\%          & 21.00\%          & 20.20\%          & 10.40\%                                                & 11.60\%                                                & 7.10\%                                                    & 19.50\%          & 27.60\%          & 20.90\%          & 12.00\%          & \textbf{29.70\%} & \textbf{26.40\%} & \textbf{11.60\%}                                       & 11.70\%                                                & \textbf{4.90\%}                                           & 18.10\%          \\ \midrule
\multicolumn{1}{c|}{\multirow{4}{*}{AdvGAN}}    & DMS        & \textbf{40.90\%} & \textbf{32.10\%} & \textbf{45.00\%} & \textbf{37.60\%} & \textbf{29.50\%} & \textbf{16.50\%}                                       & \textbf{17.50\%}                                       & \textbf{7.60\%}                                           & \textbf{28.34\%} & \textbf{46.80\%} & \textbf{31.10\%} & \textbf{6.20\%}  & \textbf{46.20\%} & \textbf{56.80\%} & \textbf{40.40\%}                                       & \textbf{42.60\%}                                       & \textbf{19.60\%}                                          & \textbf{36.21\%} \\
\multicolumn{1}{c|}{}                           & Truncating & 38.70\%          & 31.00\%          & 43.10\%          & 36.10\%          & 28.10\%          & 15.00\%                                                & 16.90\%                                                & 6.70\%                                                    & 26.95\%          & 44.10\%          & 29.50\%          & 5.70\%           & 43.90\%          & 52.80\%          & 37.50\%                                                & 38.00\%                                                & 17.60\%                                                   & 33.64\%          \\
\multicolumn{1}{c|}{}                           & Rounding   & 38.80\%          & 31.00\%          & 43.20\%          & 36.10\%          & 28.00\%          & 15.50\%                                                & 17.00\%                                                & 6.80\%                                                    & 27.05\%          & 44.30\%          & 29.40\%          & 5.70\%           & 44.00\%          & 53.00\%          & 37.30\%                                                & 37.70\%                                                & 17.40\%                                                   & 33.60\%          \\
\multicolumn{1}{c|}{}                           & Upper      & 39.00\%          & 31.00\%          & 43.10\%          & 36.00\%          & 28.10\%          & 14.90\%                                                & 16.80\%                                                & 6.70\%                                                    & 26.95\%          & 43.90\%          & 29.70\%          & 5.60\%           & 43.80\%          & 53.00\%          & 37.40\%                                                & 37.90\%                                                & 17.50\%                                                   & 33.60\%          \\ \midrule
\multicolumn{1}{c|}{\multirow{4}{*}{DI-FGSM}}   & DMS        & \textbf{57.50\%} & \textbf{52.10\%} & 42.10\%          & \textbf{39.50\%} & \textbf{41.00\%} & \textbf{18.00\%}                                       & \textbf{16.70\%}                                       & \textbf{11.60\%}                                          & \textbf{34.81\%} & \textbf{57.60\%} & \textbf{53.80\%} & \textbf{42.40\%} & \textbf{61.30\%} & \textbf{60.00\%} & \textbf{18.40\%}                                       & \textbf{17.80\%}                                       & \textbf{8.40\%}                                           & \textbf{39.96\%} \\
\multicolumn{1}{c|}{}                           & Truncating & 56.40\%          & 51.20\%          & \textbf{42.20\%} & 38.20\%          & 39.50\%          & 17.80\%                                                & 16.30\%                                                & 11.30\%                                                   & 34.11\%          & 55.90\%          & 51.90\%          & 41.60\%          & 59.50\%          & 58.20\%          & 17.90\%                                                & 17.50\%                                                & 8.20\%                                                    & 38.84\%          \\
\multicolumn{1}{c|}{}                           & Rounding   & 55.80\%          & 49.90\%          & 40.80\%          & 37.70\%          & 39.20\%          & 17.80\%                                                & 16.80\%                                                & 11.80\%                                                   & 33.73\%          & 55.10\%          & 51.30\%          & 41.10\%          & 58.50\%          & 57.30\%          & 17.90\%                                                & 17.50\%                                                & 8.60\%                                                    & 38.41\%          \\
\multicolumn{1}{c|}{}                           & Upper      & 56.60\%          & 51.50\%          & 42.10\%          & 38.40\%          & 39.60\%          & 17.70\%                                                & 16.10\%                                                & 11.30\%                                                   & 34.16\%          & 55.50\%          & 51.70\%          & 41.30\%          & 59.30\%          & 58.00\%          & 17.90\%                                                & 17.30\%                                                & 8.20\%                                                    & 38.65\%          \\ \midrule
\multicolumn{1}{c|}{\multirow{4}{*}{MI-FGSM}}   & DMS        & \textbf{61.70\%} & \textbf{55.40\%} & \textbf{49.90\%} & \textbf{45.80\%} & \textbf{44.70\%} & \textbf{21.80\%}                                       & \textbf{22.30\%}                                       & \textbf{13.30\%}                                          & \textbf{39.36\%} & \textbf{55.40\%} & \textbf{49.30\%} & \textbf{40.90\%} & \textbf{57.90\%} & \textbf{54.50\%} & \textbf{20.40\%}                                       & \textbf{20.00\%}                                       & \textbf{10.40\%}                                          & \textbf{38.60\%} \\
\multicolumn{1}{c|}{}                           & Truncating & 60.10\%          & 54.00\%          & 48.40\%          & 45.10\%          & 43.50\%          & 21.30\%                                                & 21.90\%                                                & 12.80\%                                                   & 38.39\%          & 54.60\%          & 48.30\%          & 40.00\%          & 57.20\%          & 53.00\%          & 19.90\%                                                & 19.50\%                                                & 10.30\%                                                   & 37.85\%          \\
\multicolumn{1}{c|}{}                           & Rounding   & 60.70\%          & 54.90\%          & 49.30\%          & 45.10\%          & 44.10\%          & 21.60\%                                                & 21.80\%                                                & 13.20\%                                                   & 38.84\%          & 54.80\%          & 49.00\%          & 40.60\%          & 57.60\%          & 53.70\%          & 20.00\%                                                & 19.70\%                                                & 10.20\%                                                   & 38.20\%          \\
\multicolumn{1}{c|}{}                           & Upper      & 59.90\%          & 54.10\%          & 48.20\%          & 45.20\%          & 43.30\%          & 21.30\%                                                & 21.90\%                                                & 12.80\%                                                   & 38.34\%          & 54.10\%          & 48.20\%          & 39.80\%          & 57.30\%          & 52.90\%          & 20.20\%                                                & 19.70\%                                                & 10.20\%                                                   & 37.80\%          \\ \midrule
\multicolumn{1}{c|}{\multirow{4}{*}{GE-AdvGAN}} & DMS        & \textbf{84.90\%} & \textbf{82.10\%} & \textbf{86.70\%} & \textbf{76.90\%} & \textbf{73.90\%} & \textbf{33.60\%}                                       & \textbf{39.90\%}                                       & \textbf{14.80\%}                                          & \textbf{61.60\%} & \textbf{76.10\%} & \textbf{73.50\%} & \textbf{59.60\%} & \textbf{79.90\%} & \textbf{83.10\%} & \textbf{66.40\%}                                       & \textbf{57.90\%}                                       & \textbf{50.60\%}                                          & \textbf{68.39\%} \\
\multicolumn{1}{c|}{}                           & Truncating & 83.60\%          & 80.60\%          & 85.50\%          & 75.40\%          & 72.20\%          & 32.00\%                                                & 37.20\%                                                & 13.80\%                                                   & 60.04\%          & 75.30\%          & 71.90\%          & 57.40\%          & 77.70\%          & 81.00\%          & 63.00\%                                                & 55.70\%                                                & 48.20\%                                                   & 66.28\%          \\
\multicolumn{1}{c|}{}                           & Rounding   & 83.50\%          & 80.40\%          & 85.30\%          & 75.30\%          & 71.90\%          & 31.50\%                                                & 36.60\%                                                & 13.50\%                                                   & 59.75\%          & 75.10\%          & 71.90\%          & 57.30\%          & 77.50\%          & 80.80\%          & 62.90\%                                                & 54.90\%                                                & 48.00\%                                                   & 66.05\%          \\
\multicolumn{1}{c|}{}                           & Upper      & 83.80\%          & 80.60\%          & 85.40\%          & 75.40\%          & 72.10\%          & 32.10\%                                                & 37.00\%                                                & 13.80\%                                                   & 60.03\%          & 75.10\%          & 71.90\%          & 57.50\%          & 77.50\%          & 81.00\%          & 62.80\%                                                & 55.30\%                                                & 48.10\%                                                   & 66.15\%          \\ \midrule
\multicolumn{1}{c|}{\multirow{4}{*}{SINI-FGSM}} & DMS        & \textbf{88.60\%} & \textbf{85.40\%} & \textbf{79.70\%} & \textbf{78.80\%} & \textbf{78.80\%} & \textbf{56.00\%}                                       & \textbf{50.10\%}                                       & \textbf{39.70\%}                                          & \textbf{69.64\%} & \textbf{69.30\%} & \textbf{64.80\%} & \textbf{55.60\%} & \textbf{72.40\%} & \textbf{69.80\%} & \textbf{26.30\%}                                       & 27.90\%                                                & \textbf{14.20\%}                                          & \textbf{50.04\%} \\
\multicolumn{1}{c|}{}                           & Truncating & 87.70\%          & 84.20\%          & 78.80\%          & 78.20\%          & 77.60\%          & 54.70\%                                                & 48.90\%                                                & 38.80\%                                                   & 68.61\%          & 68.10\%          & 63.10\%          & 54.00\%          & 71.70\%          & 68.00\%          & 25.50\%                                                & 27.50\%                                                & 14.00\%                                                   & 48.99\%          \\
\multicolumn{1}{c|}{}                           & Rounding   & 88.10\%          & 84.60\%          & 79.30\%          & 78.40\%          & 78.40\%          & 55.40\%                                                & 49.30\%                                                & 39.00\%                                                   & 69.06\%          & 68.70\%          & 64.10\%          & 54.50\%          & 71.90\%          & 68.60\%          & 26.00\%                                                & \textbf{28.00\%}                                       & 14.00\%                                                   & 49.48\%          \\
\multicolumn{1}{c|}{}                           & Upper      & 87.70\%          & 83.90\%          & 78.80\%          & 78.20\%          & 77.40\%          & 54.60\%                                                & 48.80\%                                                & 38.50\%                                                   & 68.49\%          & 67.80\%          & 63.10\%          & 54.10\%          & 71.50\%          & 67.80\%          & 25.40\%                                                & 27.10\%                                                & 13.90\%                                                   & 48.84\%          \\ \midrule
\multicolumn{1}{c|}{\multirow{4}{*}{NAA}}       & DMS        & \textbf{84.80\%} & \textbf{81.70\%} & \textbf{80.50\%} & \textbf{79.30\%} & \textbf{77.80\%} & \textbf{64.00\%}                                       & \textbf{58.70\%}                                       & \textbf{52.20\%}                                          & \textbf{72.38\%} & \textbf{88.60\%} & \textbf{87.20\%} & \textbf{82.70\%} & \textbf{90.40\%} & \textbf{91.30\%} & \textbf{43.00\%}                                       & \textbf{41.70\%}                                       & \textbf{23.80\%}                                          & \textbf{68.59\%} \\
\multicolumn{1}{c|}{}                           & Truncating & 84.30\%          & 81.30\%          & 79.50\%          & 78.10\%          & 76.90\%          & 63.00\%                                                & 56.40\%                                                & 50.40\%                                                   & 71.24\%          & 88.00\%          & 86.40\%          & 81.50\%          & 89.10\%          & 90.80\%          & 41.60\%                                                & 39.70\%                                                & 23.00\%                                                   & 67.51\%          \\
\multicolumn{1}{c|}{}                           & Rounding   & 84.40\%          & 81.40\%          & 79.80\%          & 78.60\%          & 77.00\%          & 63.30\%                                                & 57.30\%                                                & 50.90\%                                                   & 71.59\%          & 88.20\%          & 86.50\%          & 81.90\%          & 89.60\%          & 91.10\%          & 42.00\%                                                & 40.10\%                                                & 23.30\%                                                   & 67.84\%          \\
\multicolumn{1}{c|}{}                           & Upper      & 84.20\%          & 81.20\%          & 79.40\%          & 78.30\%          & 76.80\%          & 63.00\%                                                & 56.50\%                                                & 50.40\%                                                   & 71.23\%          & 88.00\%          & 86.30\%          & 81.30\%          & 89.20\%          & 90.40\%          & 41.20\%                                                & 39.60\%                                                & 22.80\%                                                   & 67.35\%          \\ \midrule
\multicolumn{1}{c|}{\multirow{4}{*}{MIG}}       & DMS        & \textbf{88.50\%} & \textbf{84.90\%} & \textbf{83.80\%} & \textbf{81.80\%} & \textbf{79.20\%} & \textbf{63.10\%}                                       & \textbf{58.00\%}                                       & \textbf{45.80\%}                                          & \textbf{73.14\%} & \textbf{68.20\%} & \textbf{61.70\%} & \textbf{54.10\%} & \textbf{72.50\%} & \textbf{69.70\%} & \textbf{29.50\%}                                       & \textbf{27.30\%}                                       & \textbf{15.60\%}                                          & \textbf{49.83\%} \\
\multicolumn{1}{c|}{}                           & Truncating & 88.20\%          & 84.20\%          & 83.00\%          & 80.30\%          & 77.50\%          & 61.80\%                                                & 56.40\%                                                & 44.00\%                                                   & 71.93\%          & 66.90\%          & 60.60\%          & 52.50\%          & 71.70\%          & 68.40\%          & 27.90\%                                                & 26.50\%                                                & 15.10\%                                                   & 48.70\%          \\
\multicolumn{1}{c|}{}                           & Rounding   & 88.30\%          & 84.30\%          & 83.30\%          & 81.20\%          & 78.60\%          & 62.80\%                                                & 57.20\%                                                & 45.40\%                                                   & 72.64\%          & 67.60\%          & 61.00\%          & 53.30\%          & 72.10\%          & 69.50\%          & 28.50\%                                                & 26.70\%                                                & 15.50\%                                                   & 49.28\%          \\
\multicolumn{1}{c|}{}                           & Upper      & 88.00\%          & 84.10\%          & 83.00\%          & 80.40\%          & 77.50\%          & 61.80\%                                                & 56.20\%                                                & 43.80\%                                                   & 71.85\%          & 66.60\%          & 60.30\%          & 52.50\%          & 71.70\%          & 68.30\%          & 28.00\%                                                & 26.70\%                                                & 15.00\%                                                   & 48.64\%          \\ \midrule
\multicolumn{1}{c|}{\multirow{4}{*}{DANAA}}     & DMS        & \textbf{88.80\%} & \textbf{85.90\%} & \textbf{85.40\%} & \textbf{85.30\%} & \textbf{83.40\%} & \textbf{71.70\%}                                       & \textbf{67.10\%}                                       & \textbf{57.00\%}                                          & \textbf{78.08\%} & \textbf{97.20\%} & \textbf{96.30\%} & \textbf{95.10\%} & \textbf{96.70\%} & \textbf{97.10\%} & \textbf{73.20\%}                                       & \textbf{66.10\%}                                       & \textbf{49.40\%}                                          & \textbf{83.89\%} \\
\multicolumn{1}{c|}{}                           & Truncating & 88.30\%          & 85.20\%          & 84.50\%          & 84.20\%          & 82.50\%          & 70.20\%                                                & 64.80\%                                                & 54.50\%                                                   & 76.78\%          & 96.30\%          & 95.80\%          & 94.20\%          & 96.60\%          & 96.70\%          & 70.50\%                                                & 64.00\%                                                & 46.20\%                                                   & 82.54\%          \\
\multicolumn{1}{c|}{}                           & Rounding   & 88.20\%          & 85.00\%          & 84.50\%          & 84.10\%          & 82.40\%          & 70.10\%                                                & 64.80\%                                                & 54.30\%                                                   & 76.68\%          & 96.30\%          & 95.80\%          & 94.20\%          & 96.60\%          & 96.70\%          & 70.60\%                                                & 64.30\%                                                & 45.90\%                                                   & 82.55\%          \\
\multicolumn{1}{c|}{}                           & Upper      & 88.00\%          & 85.20\%          & 84.40\%          & 84.10\%          & 82.30\%          & 70.10\%                                                & 64.50\%                                                & 54.40\%                                                   & 76.63\%          & 96.40\%          & 95.80\%          & 94.20\%          & 96.60\%          & 96.70\%          & 70.60\%                                                & 64.00\%                                                & 46.00\%                                                   & 82.54\%          \\ \midrule
\multicolumn{1}{c|}{\multirow{4}{*}{FSPS}}      & DMS        & \textbf{89.00\%} & \textbf{88.20\%} & \textbf{86.80\%} & \textbf{84.90\%} & \textbf{86.00\%} & \textbf{82.50\%}                                       & \textbf{80.40\%}                                       & \textbf{79.80\%}                                          & \textbf{84.70\%} & \textbf{94.80\%} & \textbf{94.40\%} & \textbf{93.40\%} & \textbf{95.70\%} & \textbf{94.70\%} & \textbf{86.50\%}                                       & \textbf{85.30\%}                                       & \textbf{81.70\%}                                          & \textbf{90.81\%} \\
\multicolumn{1}{c|}{}                           & Truncating & 88.50\%          & 87.70\%          & 85.90\%          & 84.20\%          & 85.30\%          & 81.50\%                                                & 79.50\%                                                & 78.60\%                                                   & 83.90\%          & 94.30\%          & 93.90\%          & 92.40\%          & 95.20\%          & 94.20\%          & 85.60\%                                                & 85.00\%                                                & 79.50\%                                                   & 90.01\%          \\
\multicolumn{1}{c|}{}                           & Rounding   & 88.60\%          & 87.70\%          & 86.50\%          & 84.60\%          & 85.50\%          & 81.50\%                                                & 79.70\%                                                & 78.80\%                                                   & 84.11\%          & 94.60\%          & 93.90\%          & 92.90\%          & 95.40\%          & 94.40\%          & 85.90\%                                                & 85.00\%                                                & 80.10\%                                                   & 90.28\%          \\
\multicolumn{1}{c|}{}                           & Upper      & 88.50\%          & 87.70\%          & 86.10\%          & 84.20\%          & 85.10\%          & 81.50\%                                                & 79.50\%                                                & 78.60\%                                                   & 83.90\%          & 94.30\%          & 93.80\%          & 92.30\%          & 95.20\%          & 94.10\%          & 85.50\%                                                & 84.90\%                                                & 79.50\%                                                   & 89.95\%          \\ \midrule
\multicolumn{1}{c|}{\multirow{4}{*}{SSA}}       & DMS        & \textbf{90.50\%} & \textbf{90.60\%} & \textbf{86.10\%} & \textbf{85.40\%} & \textbf{86.30\%} & \textbf{81.00\%}                                       & \textbf{78.00\%}                                       & \textbf{77.70\%}                                          & \textbf{84.45\%} & \textbf{90.60\%} & \textbf{90.20\%} & \textbf{87.80\%} & \textbf{93.10\%} & \textbf{92.10\%} & \textbf{80.30\%}                                       & \textbf{78.00\%}                                       & \textbf{70.50\%}                                          & \textbf{85.33\%} \\
\multicolumn{1}{c|}{}                           & Truncating & 89.80\%          & 90.00\%          & 85.60\%          & 84.60\%          & 85.00\%          & 79.80\%                                                & 77.10\%                                                & 76.90\%                                                   & 83.60\%          & 89.80\%          & 89.90\%          & 86.80\%          & 92.80\%          & 91.70\%          & 79.30\%                                                & 77.00\%                                                & 69.00\%                                                   & 84.54\%          \\
\multicolumn{1}{c|}{}                           & Rounding   & 89.90\%          & 90.30\%          & 85.70\%          & 85.10\%          & 85.50\%          & 79.80\%                                                & 77.40\%                                                & 77.00\%                                                   & 83.84\%          & 90.00\%          & 89.90\%          & 87.10\%          & 92.90\%          & 91.90\%          & 79.60\%                                                & 77.30\%                                                & 69.30\%                                                   & 84.75\%          \\
\multicolumn{1}{c|}{}                           & Upper      & 89.80\%          & 90.00\%          & 85.50\%          & 84.50\%          & 84.90\%          & 79.80\%                                                & 76.90\%                                                & 76.60\%                                                   & 83.50\%          & 89.80\%          & 89.90\%          & 86.80\%          & 92.80\%          & 91.50\%          & 79.30\%                                                & 77.20\%                                                & 68.90\%                                                   & 84.53\%          \\ \midrule
\multicolumn{1}{c|}{\multirow{4}{*}{GRA}}      & DMS        & \textbf{88.20\%}               & \textbf{86.90\%}               & \textbf{83.10\%}                  & \textbf{83.00\%}                   & \textbf{81.80\%}                   & \textbf{73.30\%}                    & \textbf{69.20\%}                    & \textbf{67.60\%}                       & \textbf{79.14\%} & \textbf{89.20\%} & \textbf{87.50\%}               & \textbf{86.00\%}                   & \textbf{92.90\%}                  & \textbf{91.80\%}                   & \textbf{69.00\%}                    & \textbf{67.90\%}                    & \textbf{57.10\%}                       & \textbf{80.36\%} \\
\multicolumn{1}{c|}{}      & Truncating & 87.50\%                        & 86.00\%                        & 81.70\%                           & 81.90\%                            & 81.40\%                            & 71.70\%                             & 68.10\%                             & 65.50\%                                & 77.98\%          & 88.90\%          & 86.80\%                        & 84.80\%                            & 92.00\%                           & 91.10\%                            & 68.30\%                             & 66.60\%                             & 56.10\%                                & 79.56\%          \\
\multicolumn{1}{c|}{}      & Rounding   & 87.60\%                        & 86.30\%                        & 82.00\%                           & 82.30\%                            & 81.40\%                            & 72.20\%                             & 68.70\%                             & 66.20\%                                & 78.34\%          & 88.90\%          & 87.20\%                        & 85.30\%                            & 92.60\%                           & 91.50\%                            & 68.60\%                             & 66.40\%                             & 56.50\%                                & 79.82\%          \\
\multicolumn{1}{c|}{} & Upper      & 87.40\%                        & 86.00\%                        & 81.80\%                           & 81.80\%                            & 81.30\%                            & 71.70\%                             & 68.00\%                             & 65.70\%                                & 77.96\%          & 88.80\%          & 86.60\%                        & 84.70\%                            & 91.80\%                           & 91.00\%                            & 68.30\%                             & 66.50\%                             & 55.90\%                                & 79.43\%          \\ \bottomrule
\end{tabular}%
}
\end{table*}

% \usepackage{multirow}

\subsection{Ablation Study}

\subsubsection{The Impact of Different Learning Rates}
To validate the performance of each method with different learning rates, we choose five learning rates of 0.001, 0.005, 0.01, 0.015, and 0.02. The corresponding dataset used is CIFAR-100, the model is VGG16-BN, and the attack method is C\&W. The early stop strategy is used, and the threshold and the number of iterations are 1 and 200, respectively.

As depicted in the line graph of the learning rate in Figure.~\ref{fig:Attacksuccessrate}, it is apparent that the DMS method consistently outperforms the other methods across a range of learning rates. Spanning from 0.001 to 0.02, DMS consistently attains the highest attack success rate, escalating from 0.9534 to 0.9591. The DMS-AI method also exhibits relatively good performance, following the performance of DMS.

Conversely, the Rounding method demonstrates the lowest success rate among all the methods. As the learning rates increase, we can see a gradual improvement in the success rate, which eventually stabilizes at approximately 0.4. Furthermore, while the remaining methods may outperform the Rounding method in terms of attack success rate, these alternatives have exhibited a decrease in performance as the learning rate is elevated. The DMS method, demonstrates a more robust and higher performance in terms of the success attack rate, regardless of the learning rate changes.

\subsubsection{The Impact of Different Thresholds (Maximum perturbations)}
To validate the performance of each method under different thresholds, we choose ten thresholds of $0.5, 1, 1.5, 2, 2.5, 3, 3.5, 4, 4.5$ and $5$. The evaluation dataset is CIFAR-100, the model is VGG16-BN, and the attack method is C\&W. Using the early stop strategy, the learning rate and the number of iterations are 0.01 and 100, respectively.

\begin{figure*}[htbp]
  \centering
  % \vspace{-3cm}
  % \includegraphics[scale=0.11]{images/diff_lr.pdf}
  \includegraphics[width=\textwidth]{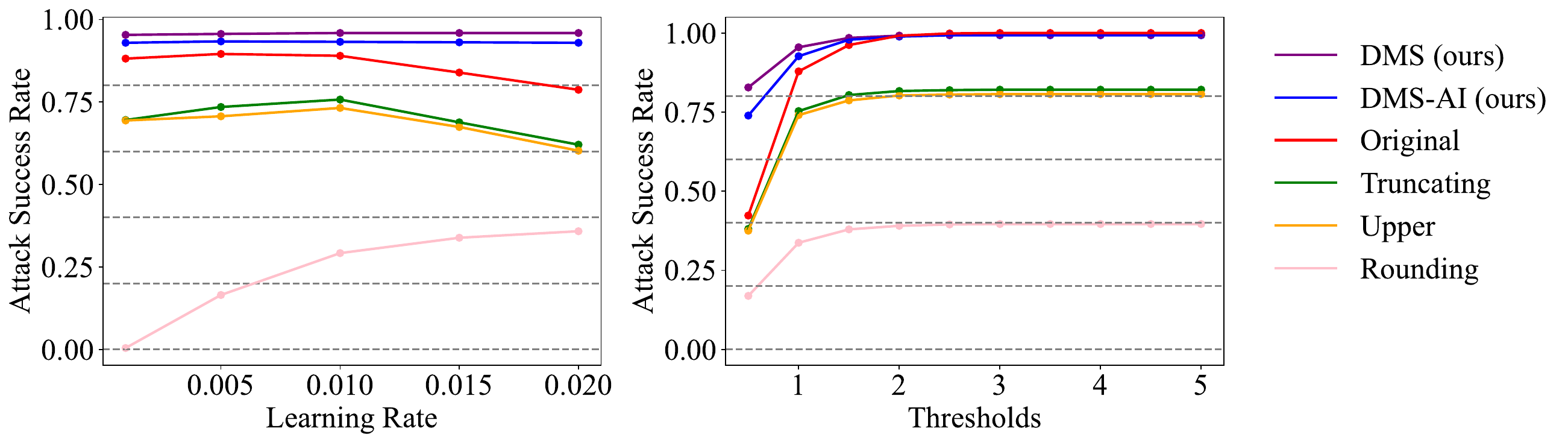}
  \caption{Attack success rate under different learning rates and thresholds (maximum perturbations)}
  \label{fig:Attacksuccessrate}
\end{figure*}

From the thresholds line graph in Figure.~\ref{fig:Attacksuccessrate}, it is evident that the DMS method consistently outperforms other methods across various threshold values. Moreover, the success rate of the DMS method increases as the threshold value increases. For instance, at a threshold of 0.5, the success rate improves from 0.8280 to 0.9944. The DMS-AI method also achieves relatively good performance, with the success rate increasing from 0.7391 to 0.9929 as the threshold increases from 0.5 to 5. On the contrary, the Rounding method exhibits relatively inferior performance, with only slight improvements as the threshold increases, significantly lower compared to other methods. Specifically, when the threshold increases from 0.5 to 5, the success rate of the Rounding method only increases from 0.1693 to 0.3963. Furthermore, while the threshold shows a positive relationship with the attack success rate, it becomes stable when the threshold value is larger than two. In summary, different threshold values may have various effects on the attack success rates for different methods, yet the DMS method can consistently perform the best regardless of the threshold value.

\begin{figure*}
    \centering
    \includegraphics[width=\linewidth]{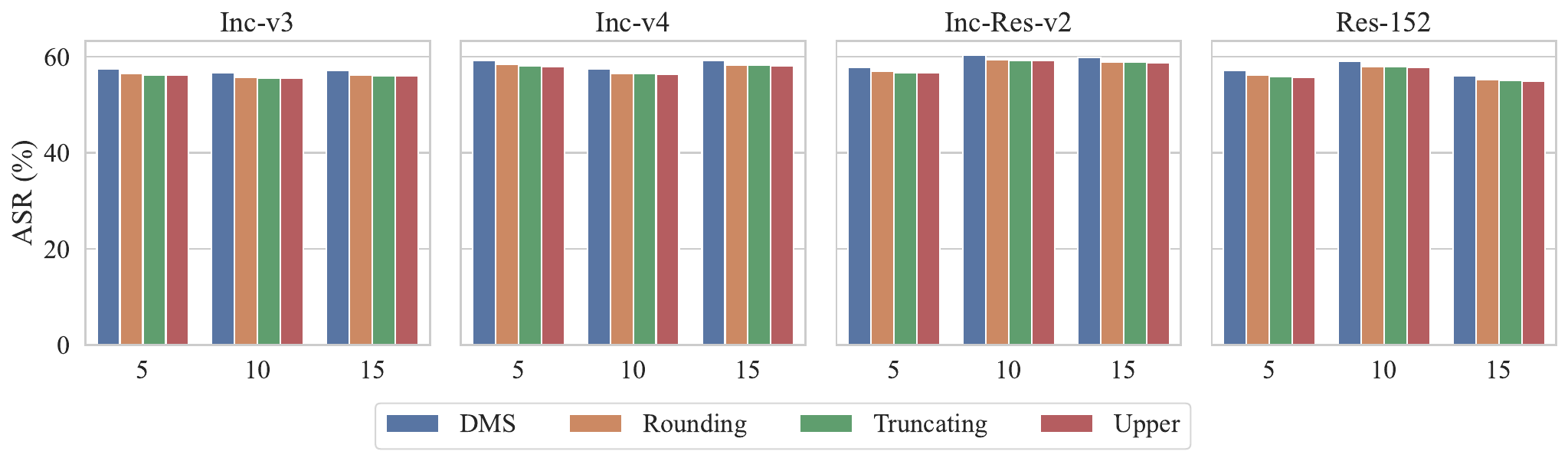}
    \caption{Transferable Performance of Various Methods by Steps on Models: This figure displays how different methods, like DMS and Rounding, perform across varying steps on surrogate models Inc-v3, Inc-v4, Inc-Res-v2, and Res-152.}
    \label{fig:diff-step}
\end{figure*}

\begin{figure}[htbp]
  \centering
  % \vspace{-3cm}
  % \includegraphics[scale=0.11]{diff_lr.pdf}
  \includegraphics[width=\linewidth]{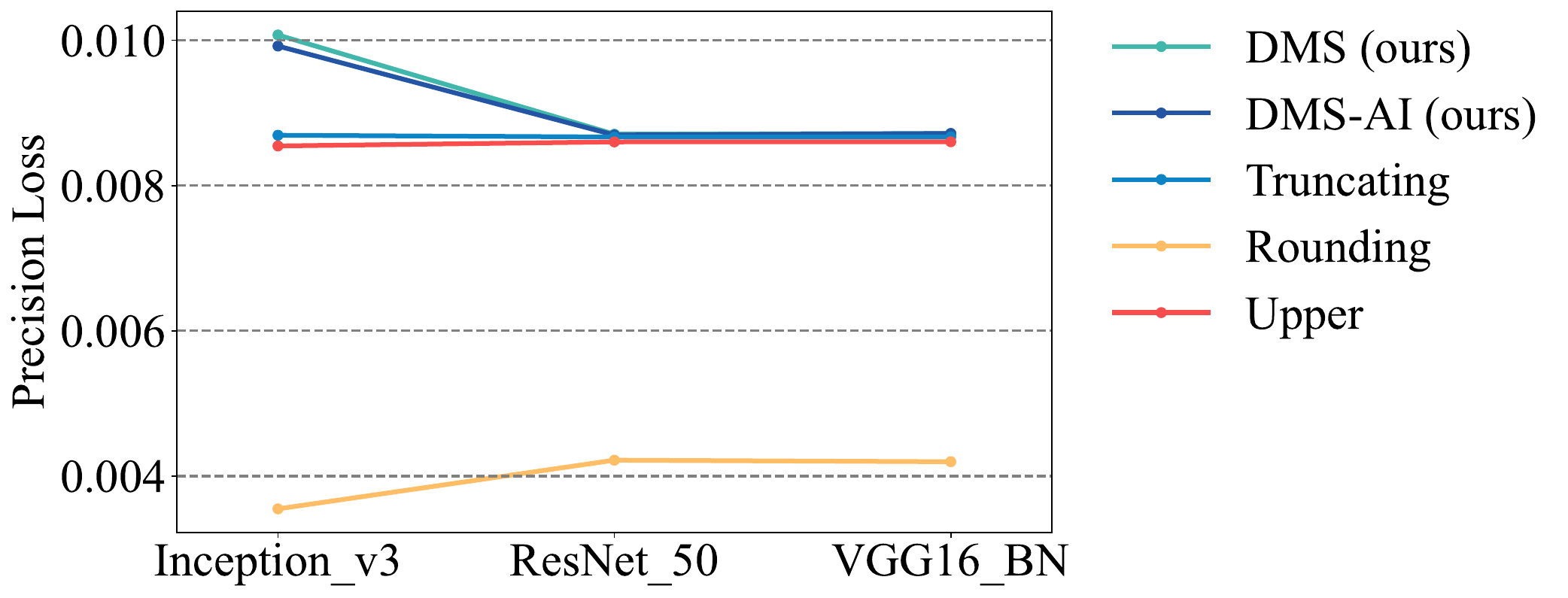}
  \caption{Precision loss of different models on the ImageNet dataset}
  \label{fig:imgerror}
\end{figure}

\subsubsection{Precision Loss under Different Models}
To evaluate the impact of integerizing on precision under various models, we select three models for our investigation: Inception-v3, ResNet-50, and VGG16-BN. The ImageNet dataset is employed, and the C\&W attack method is applied. Our experimental settings included a learning rate of 0.01, a threshold of 0.3, and 100 iterations. To ensure optimal results, we have implemented the early stopping strategy. Meanwhile, we give the definition of precision loss to compare the corresponding variation of each method.
\begin{equation}
Precision \ loss=mean( \left | P'-P \right |) 
\end{equation}

Figure.~\ref{fig:imgerror} reveals that both the DMS and DMS-AI methods have minimum precision loss in comparison to the standardised approaches across various models, in particularly for ResNet-50 and VGG16-BN. Remarkably, even in the Inception-v3 model, the precision loss is merely 0.13\% when comparing with the truncating operation. Thus, we consider the overall performance of DMS has been largely optimized to handle the information loss in image processing for adversarial samples.

\subsubsection{Transferable ASR under Different Steps}

In this section, we conducted experiments on four methods: DMS, Rounding, Truncating, and Upper, using three different step counts: 5, 10, and 15. As depicted in Figure~\ref{fig:diff-step}, when using Inception-v3 (Inc-v3) and Inception-v4 (Inc-v4) as surrogate models, the optimal step count for transfer attacks is 5. However, when using Inception-ResNet-v2 (Inc-Res-v2) and ResNet-152 (Res-152) as surrogate models, the optimal step count becomes 10. Overall, DMS consistently outperforms the other three methods. For more detailed experimental data, please refer to the Appendix and our GitHub repository.

\subsection{Time Cost}
For the time cost of DMS-AI and DMS-AS steps in our algorithm, in Section 3.2, we can see DMS-AI exclusively performs adversarial integerizing on the pixel values in the RGB image according to the gradient direction once. This step involves calculating gradients for individual pixels for only one instance. Regarding DMS-AS, it is essentially based on the gradient attribution of Integrated Gradient~\cite{sundararajan2017axiomatic}. Since we are implementing IG as the basis for DMS-AS, we find that the computational efficiency is consistent, as observed in our experiments. In summary, the increased time complexity of the DMS algorithm is minimal, yet it proves that DMS can effectively address the attack failure concern due to the precision loss challenge in image processing for file formats.

\section{Conclusion}
In this paper, we present Do More Steps (DMS) algorithm aimed to address the impact of information loss during image processing for adversarial samples, which loss can largely decrease the attack success rate. Specifically, we first employ adversarial integerization to ensure that the pixel point is integerized towards a meaningful direction for an adversarial attack. Secondly, considering the problem of gradient oscillation and latent attack failure may exist after adversarial integerization, we integrate the attribution selection method based on an integrated gradient-based attribution algorithm to select and truncate the pixels with the best attribution results in the adjusted pixel space to improve the attack success rate. Furthermore, we perform a comprehensive analysis of the precision loss. Extensive experiment results corroborate the superiority of our method for retaining the attack success rates. Moreover, we statistically analyse the precision loss to compare with other integerization approaches. It is noted that the precision loss of our DMS algorithm is expected to be minimal, while simultaneously ensuring superior performance of attack success rates. We anticipate this work will shed the lights on future computer engineering research on pragmatic adversarial attacks in machine learning and artificial intelligence areas.

% \section{Acknowledgments}
% Identification of funding sources and other support, and thanks to
% individuals and groups that assisted in the research and the
% preparation of the work should be included in an acknowledgment
% section, which is placed just before the reference section in your
% document.

% %%
% %% The acknowledgments section is defined using the "acks" environment
% %% (and NOT an unnumbered section). This ensures the proper
% %% identification of the section in the article metadata, and the
% %% consistent spelling of the heading.
% \begin{acks}
% \end{acks}

\balance
\bibliographystyle{ACM-Reference-Format}
\bibliography{ref}

\newpage
\appendix

\section*{Appendix: More Experimental Results}
In the Appendix, we provide additional parameter ablation experiment results, including tests using Inception-v3 and Inception-v4 as surrogate models, with the number of attack steps set to 5 and 10, respectively. It is evident from these experiments that the DMS method enhances the attack capability of the adversarial methods in all scenarios. Regardless of the model, the type of attack method, or the parameter settings of the attack method, the DMS approach consistently and universally achieves an average increase of approximately 1\% in Attack Success Rate (ASR). This thoroughly demonstrates the universality of our DMS method. 

More experimental data can be found in our GitHub repository.

\begin{table*}[h!]
\centering
\caption{Comparative Analysis of Attack Success Rates using Inception-v3 as Surrogate Model. This table presents a comprehensive evaluation of various attack methods across multiple models.}
\label{aptab:Inc-v3}
\resizebox{\textwidth}{!}{%
\begin{tabular}{@{}cc|ccccccccc|ccccccccc@{}}
\toprule
\multicolumn{2}{c|}{Step}                                    & \multicolumn{9}{c|}{5}                                                                                                 & \multicolumn{9}{c}{15}                                                                                                 \\ \midrule
\multicolumn{1}{c|}{Attack}                     & Method     & Inc-v4           & Inc-Res-v2           & Res-50           & Res-101          & Res-152          & \begin{tabular}[c]{@{}c@{}}Inc-v3\\ -ens3\end{tabular} & \begin{tabular}[c]{@{}c@{}}Inc-v3\\ -ens4\end{tabular} & \begin{tabular}[c]{@{}c@{}}Inc-Res\\ -v2-ens\end{tabular} & Average          & Inc-v4           & Inc-Res-v2           & Res-50           & Res-101          & Res-152          & \begin{tabular}[c]{@{}c@{}}Inc-v3\\ -ens3\end{tabular} & \begin{tabular}[c]{@{}c@{}}Inc-v3\\ -ens4\end{tabular} & \begin{tabular}[c]{@{}c@{}}Inc-Res\\ -v2-ens\end{tabular} & Average          \\ \midrule
\multicolumn{1}{c|}{\multirow{4}{*}{AdvGAN}}    & DMS        & 63.60\%      & 37.00\%    & 68.40\%   & 72.60\%    & 60.20\%    & 50.70\%     & 55.70\%     & 28.50\%        & 54.59\% & 50.70\%      & 34.00\%    & 69.70\%   & 67.00\%    & 55.00\%    & 46.30\%     & 46.10\%     & 13.60\%        & 47.80\% \\
\multicolumn{1}{c|}{}                           & Truncating & 60.00\%      & 34.50\%    & 65.60\%   & 69.90\%    & 57.10\%    & 47.40\%     & 52.30\%     & 26.20\%        & 51.63\% & 47.30\%      & 32.00\%    & 65.90\%   & 63.70\%    & 51.70\%    & 43.20\%     & 43.20\%     & 12.50\%        & 44.94\% \\
\multicolumn{1}{c|}{}                           & Rounding   & 59.50\%      & 34.70\%    & 65.70\%   & 69.90\%    & 57.10\%    & 47.40\%     & 51.80\%     & 26.50\%        & 51.58\% & 47.50\%      & 32.20\%    & 65.70\%   & 63.80\%    & 51.50\%    & 42.70\%     & 43.10\%     & 12.50\%        & 44.88\% \\
\multicolumn{1}{c|}{}                           & Upper      & 59.40\%      & 34.60\%    & 65.70\%   & 69.80\%    & 56.90\%    & 47.50\%     & 52.10\%     & 26.20\%        & 51.53\% & 47.10\%      & 31.90\%    & 65.70\%   & 63.80\%    & 51.60\%    & 43.20\%     & 43.00\%     & 12.40\%        & 44.84\% \\ \midrule
\multicolumn{1}{c|}{\multirow{4}{*}{DANAA}}     & DMS        & 89.50\%      & 87.80\%    & 87.20\%   & 85.90\%    & 84.50\%    & 62.20\%     & 57.70\%     & 39.60\%        & 74.30\% & 94.10\%      & 92.10\%    & 91.70\%   & 91.90\%    & 90.50\%    & 76.10\%     & 72.00\%     & 55.30\%        & 82.96\% \\
\multicolumn{1}{c|}{}                           & Truncating & 88.80\%      & 87.00\%    & 85.80\%   & 84.80\%    & 83.00\%    & 59.60\%     & 55.90\%     & 37.40\%        & 72.79\% & 93.80\%      & 91.60\%    & 91.00\%   & 91.60\%    & 90.10\%    & 74.40\%     & 69.70\%     & 52.70\%        & 81.86\% \\
\multicolumn{1}{c|}{}                           & Rounding   & 88.60\%      & 86.80\%    & 85.40\%   & 84.50\%    & 82.80\%    & 59.20\%     & 56.00\%     & 37.10\%        & 72.55\% & 93.80\%      & 91.60\%    & 91.00\%   & 91.50\%    & 90.00\%    & 74.00\%     & 70.00\%     & 52.30\%        & 81.78\% \\
\multicolumn{1}{c|}{}                           & Upper      & 88.70\%      & 87.00\%    & 85.80\%   & 84.80\%    & 83.00\%    & 59.50\%     & 56.40\%     & 37.40\%        & 72.83\% & 93.80\%      & 91.60\%    & 91.00\%   & 91.70\%    & 90.20\%    & 74.60\%     & 69.70\%     & 52.60\%        & 81.90\% \\ \midrule
\multicolumn{1}{c|}{\multirow{4}{*}{DI-FGSM}}   & DMS        & 53.10\%      & 46.60\%    & 43.50\%   & 41.10\%    & 40.50\%    & 18.70\%     & 19.10\%     & 9.60\%         & 34.03\% & 44.80\%      & 36.80\%    & 36.60\%   & 33.60\%    & 31.50\%    & 17.50\%     & 18.30\%     & 7.80\%         & 28.36\% \\
\multicolumn{1}{c|}{}                           & Truncating & 52.00\%      & 44.90\%    & 42.90\%   & 39.50\%    & 39.20\%    & 18.60\%     & 18.80\%     & 9.60\%         & 33.19\% & 43.30\%      & 36.50\%    & 36.20\%   & 32.70\%    & 30.50\%    & 17.30\%     & 17.80\%     & 7.80\%         & 27.76\% \\
\multicolumn{1}{c|}{}                           & Rounding   & 52.50\%      & 45.50\%    & 43.50\%   & 39.80\%    & 40.10\%    & 19.00\%     & 19.20\%     & 10.10\%        & 33.71\% & 40.30\%      & 34.10\%    & 34.50\%   & 31.00\%    & 28.70\%    & 16.60\%     & 17.50\%     & 7.40\%         & 26.26\% \\
\multicolumn{1}{c|}{}                           & Upper      & 52.00\%      & 44.80\%    & 43.00\%   & 39.70\%    & 39.00\%    & 18.30\%     & 18.80\%     & 9.60\%         & 33.15\% & 43.30\%      & 36.80\%    & 36.00\%   & 32.50\%    & 30.30\%    & 17.20\%     & 17.70\%     & 7.80\%         & 27.70\% \\ \midrule
\multicolumn{1}{c|}{\multirow{4}{*}{FSPS}}      & DMS        & 89.50\%      & 88.40\%    & 84.70\%   & 85.10\%    & 85.20\%    & 79.50\%     & 78.00\%     & 67.80\%        & 82.28\% & 92.80\%      & 91.10\%    & 89.20\%   & 89.50\%    & 88.30\%    & 85.20\%     & 83.80\%     & 75.80\%        & 86.96\% \\
\multicolumn{1}{c|}{}                           & Truncating & 88.20\%      & 87.80\%    & 83.90\%   & 84.10\%    & 84.40\%    & 77.90\%     & 76.50\%     & 66.10\%        & 81.11\% & 92.70\%      & 90.50\%    & 88.40\%   & 88.80\%    & 87.70\%    & 84.50\%     & 82.80\%     & 73.90\%        & 86.16\% \\
\multicolumn{1}{c|}{}                           & Rounding   & 88.50\%      & 88.20\%    & 84.30\%   & 84.50\%    & 84.50\%    & 78.60\%     & 77.10\%     & 66.60\%        & 81.54\% & 92.70\%      & 90.90\%    & 88.70\%   & 89.20\%    & 88.20\%    & 84.80\%     & 83.90\%     & 75.20\%        & 86.70\% \\
\multicolumn{1}{c|}{}                           & Upper      & 88.20\%      & 87.90\%    & 83.70\%   & 83.80\%    & 84.40\%    & 78.00\%     & 76.70\%     & 66.10\%        & 81.10\% & 92.70\%      & 90.50\%    & 88.50\%   & 88.80\%    & 87.70\%    & 84.30\%     & 82.70\%     & 73.70\%        & 86.11\% \\ \midrule
\multicolumn{1}{c|}{\multirow{4}{*}{GE-AdvGAN}} & DMS        & 70.60\%      & 49.00\%    & 78.10\%   & 82.90\%    & 73.50\%    & 33.80\%     & 55.40\%     & 29.80\%        & 59.14\% & 77.60\%      & 63.30\%    & 85.00\%   & 89.30\%    & 80.60\%    & 59.50\%     & 62.70\%     & 46.40\%        & 70.55\% \\
\multicolumn{1}{c|}{}                           & Truncating & 69.00\%      & 47.50\%    & 76.50\%   & 81.60\%    & 71.70\%    & 31.90\%     & 53.70\%     & 28.40\%        & 57.54\% & 76.60\%      & 61.20\%    & 84.00\%   & 87.60\%    & 78.80\%    & 56.00\%     & 59.30\%     & 44.00\%        & 68.44\% \\
\multicolumn{1}{c|}{}                           & Rounding   & 69.20\%      & 47.50\%    & 76.50\%   & 81.30\%    & 71.40\%    & 31.40\%     & 53.70\%     & 28.40\%        & 57.43\% & 76.40\%      & 61.00\%    & 83.90\%   & 87.50\%    & 78.80\%    & 56.10\%     & 59.00\%     & 43.60\%        & 68.29\% \\
\multicolumn{1}{c|}{}                           & Upper      & 69.10\%      & 47.50\%    & 76.30\%   & 81.60\%    & 71.70\%    & 31.90\%     & 53.90\%     & 28.00\%        & 57.50\% & 76.60\%      & 61.20\%    & 84.00\%   & 87.70\%    & 78.70\%    & 55.90\%     & 59.30\%     & 44.00\%        & 68.43\% \\ \midrule
\multicolumn{1}{c|}{\multirow{4}{*}{I-FGSM}}    & DMS        & 40.40\%      & 33.50\%    & 33.90\%   & 31.90\%    & 28.40\%    & 14.90\%     & 13.00\%     & 7.10\%         & 25.39\% & 22.90\%      & 14.30\%    & 20.90\%   & 17.70\%    & 15.90\%    & 11.20\%     & 11.20\%     & 4.70\%         & 14.85\% \\
\multicolumn{1}{c|}{}                           & Truncating & 39.10\%      & 32.40\%    & 34.20\%   & 31.10\%    & 27.90\%    & 14.70\%     & 13.20\%     & 7.10\%         & 24.96\% & 22.90\%      & 14.60\%    & 21.00\%   & 17.60\%    & 15.20\%    & 10.60\%     & 10.70\%     & 4.60\%         & 14.65\% \\
\multicolumn{1}{c|}{}                           & Rounding   & 39.80\%      & 32.90\%    & 34.60\%   & 31.20\%    & 28.30\%    & 14.60\%     & 13.80\%     & 7.70\%         & 25.36\% & 20.10\%      & 12.50\%    & 19.10\%   & 16.00\%    & 13.90\%    & 10.20\%     & 11.10\%     & 4.70\%         & 13.45\% \\
\multicolumn{1}{c|}{}                           & Upper      & 39.00\%      & 32.40\%    & 34.10\%   & 30.70\%    & 28.10\%    & 14.60\%     & 13.10\%     & 7.20\%         & 24.90\% & 23.00\%      & 14.50\%    & 20.80\%   & 17.70\%    & 15.30\%    & 11.00\%     & 10.90\%     & 4.60\%         & 14.73\% \\ \midrule
\multicolumn{1}{c|}{\multirow{4}{*}{MI-FGSM}}   & DMS        & 56.90\%      & 53.80\%    & 51.20\%   & 48.70\%    & 45.00\%    & 23.60\%     & 22.90\%     & 12.00\%        & 39.26\% & 47.50\%      & 45.20\%    & 45.00\%   & 40.20\%    & 38.60\%    & 21.60\%     & 21.20\%     & 10.20\%        & 33.69\% \\
\multicolumn{1}{c|}{}                           & Truncating & 55.90\%      & 52.50\%    & 50.40\%   & 47.50\%    & 43.90\%    & 23.70\%     & 22.40\%     & 11.50\%        & 38.48\% & 46.90\%      & 44.80\%    & 44.10\%   & 39.60\%    & 38.00\%    & 21.70\%     & 21.00\%     & 9.90\%         & 33.25\% \\
\multicolumn{1}{c|}{}                           & Rounding   & 56.80\%      & 52.80\%    & 51.00\%   & 47.90\%    & 44.80\%    & 23.50\%     & 22.60\%     & 12.10\%        & 38.94\% & 47.40\%      & 45.20\%    & 44.30\%   & 40.20\%    & 38.40\%    & 21.70\%     & 21.20\%     & 10.70\%        & 33.64\% \\
\multicolumn{1}{c|}{}                           & Upper      & 56.10\%      & 52.30\%    & 50.20\%   & 47.40\%    & 43.80\%    & 23.60\%     & 22.30\%     & 11.50\%        & 38.40\% & 46.80\%      & 44.60\%    & 43.90\%   & 39.40\%    & 37.70\%    & 21.70\%     & 21.00\%     & 9.90\%         & 33.13\% \\ \midrule
\multicolumn{1}{c|}{\multirow{4}{*}{MIG}}       & DMS        & 74.70\%      & 73.50\%    & 72.40\%   & 68.20\%    & 67.90\%    & 41.80\%     & 40.30\%     & 23.80\%        & 57.83\% & 69.30\%      & 68.10\%    & 69.00\%   & 64.30\%    & 62.70\%    & 37.90\%     & 39.20\%     & 20.70\%        & 53.90\% \\
\multicolumn{1}{c|}{}                           & Truncating & 73.40\%      & 72.80\%    & 71.20\%   & 66.40\%    & 66.80\%    & 40.30\%     & 39.40\%     & 23.20\%        & 56.69\% & 68.40\%      & 66.60\%    & 68.00\%   & 62.90\%    & 61.80\%    & 37.20\%     & 38.20\%     & 20.10\%        & 52.90\% \\
\multicolumn{1}{c|}{}                           & Rounding   & 74.00\%      & 73.10\%    & 71.70\%   & 67.10\%    & 67.50\%    & 40.90\%     & 39.30\%     & 23.40\%        & 57.13\% & 69.20\%      & 68.10\%    & 69.00\%   & 64.30\%    & 62.30\%    & 37.80\%     & 39.10\%     & 20.80\%        & 53.83\% \\
\multicolumn{1}{c|}{}                           & Upper      & 73.60\%      & 72.50\%    & 71.30\%   & 66.30\%    & 66.80\%    & 40.50\%     & 39.20\%     & 23.00\%        & 56.65\% & 68.20\%      & 66.70\%    & 68.00\%   & 62.80\%    & 61.40\%    & 37.10\%     & 37.90\%     & 19.80\%        & 52.74\% \\ \midrule
\multicolumn{1}{c|}{\multirow{4}{*}{NAA}}       & DMS        & 87.10\%      & 86.50\%    & 83.10\%   & 82.00\%    & 80.90\%    & 54.60\%     & 53.50\%     & 33.00\%        & 70.09\% & 88.30\%      & 87.20\%    & 84.20\%   & 83.90\%    & 82.60\%    & 59.60\%     & 57.40\%     & 36.20\%        & 72.43\% \\
\multicolumn{1}{c|}{}                           & Truncating & 86.10\%      & 85.60\%    & 81.80\%   & 81.30\%    & 80.10\%    & 53.30\%     & 51.70\%     & 31.90\%        & 68.98\% & 87.70\%      & 86.00\%    & 83.30\%   & 82.70\%    & 81.20\%    & 57.10\%     & 55.40\%     & 35.10\%        & 71.06\% \\
\multicolumn{1}{c|}{}                           & Rounding   & 86.50\%      & 85.90\%    & 82.20\%   & 81.50\%    & 80.40\%    & 53.60\%     & 51.90\%     & 32.20\%        & 69.28\% & 88.00\%      & 86.50\%    & 83.70\%   & 83.20\%    & 82.50\%    & 58.50\%     & 56.80\%     & 35.30\%        & 71.81\% \\
\multicolumn{1}{c|}{}                           & Upper      & 86.00\%      & 85.50\%    & 82.00\%   & 81.10\%    & 80.20\%    & 53.30\%     & 51.40\%     & 31.90\%        & 68.93\% & 87.60\%      & 85.90\%    & 83.40\%   & 82.70\%    & 81.10\%    & 56.90\%     & 55.20\%     & 35.10\%        & 70.99\% \\ \midrule
\multicolumn{1}{c|}{\multirow{4}{*}{SINI-FGSM}} & DMS        & 76.00\%      & 77.30\%    & 74.10\%   & 71.20\%    & 69.90\%    & 38.90\%     & 39.40\%     & 23.60\%        & 58.80\% & 78.30\%      & 75.30\%    & 72.90\%   & 68.70\%    & 67.50\%    & 38.10\%     & 39.40\%     & 22.80\%        & 57.88\% \\
\multicolumn{1}{c|}{}                           & Truncating & 75.20\%      & 76.30\%    & 73.00\%   & 69.50\%    & 67.90\%    & 38.30\%     & 38.60\%     & 23.00\%        & 57.73\% & 77.10\%      & 74.20\%    & 71.80\%   & 67.60\%    & 66.10\%    & 37.40\%     & 38.30\%     & 22.10\%        & 56.83\% \\
\multicolumn{1}{c|}{}                           & Rounding   & 75.60\%      & 77.00\%    & 73.70\%   & 71.00\%    & 69.20\%    & 38.60\%     & 38.70\%     & 23.50\%        & 58.41\% & 78.20\%      & 75.10\%    & 73.00\%   & 68.60\%    & 67.10\%    & 38.50\%     & 39.20\%     & 22.60\%        & 57.79\% \\
\multicolumn{1}{c|}{}                           & Upper      & 75.10\%      & 76.30\%    & 72.90\%   & 69.60\%    & 67.90\%    & 38.20\%     & 38.20\%     & 23.20\%        & 57.68\% & 76.70\%      & 73.80\%    & 72.10\%   & 67.40\%    & 66.00\%    & 37.40\%     & 38.20\%     & 22.30\%        & 56.74\% \\ \midrule
\multicolumn{1}{c|}{\multirow{4}{*}{SSA}}       & DMS        & 87.30\%      & 85.70\%    & 81.30\%   & 78.90\%    & 79.60\%    & 71.40\%     & 70.90\%     & 55.20\%        & 76.29\% & 89.20\%      & 87.40\%    & 83.60\%   & 82.40\%    & 81.90\%    & 76.60\%     & 76.80\%     & 62.80\%        & 80.09\% \\
\multicolumn{1}{c|}{}                           & Truncating & 86.30\%      & 84.60\%    & 80.30\%   & 78.00\%    & 78.40\%    & 70.00\%     & 69.50\%     & 53.80\%        & 75.11\% & 88.30\%      & 86.50\%    & 82.60\%   & 81.60\%    & 81.20\%    & 76.20\%     & 76.30\%     & 61.70\%        & 79.30\% \\
\multicolumn{1}{c|}{}                           & Rounding   & 86.60\%      & 85.30\%    & 81.10\%   & 77.80\%    & 79.00\%    & 71.00\%     & 70.20\%     & 54.00\%        & 75.63\% & 88.70\%      & 87.10\%    & 83.40\%   & 82.30\%    & 81.60\%    & 77.00\%     & 76.60\%     & 62.40\%        & 79.89\% \\
\multicolumn{1}{c|}{}                           & Upper      & 86.20\%      & 84.60\%    & 80.40\%   & 77.80\%    & 78.40\%    & 70.00\%     & 69.40\%     & 53.50\%        & 75.04\% & 88.30\%      & 86.40\%    & 82.60\%   & 81.60\%    & 81.30\%    & 76.10\%     & 76.20\%     & 61.70\%        & 79.28\% \\ \midrule
\multicolumn{1}{c|}{\multirow{4}{*}{GRA}} & DMS        & 86.50\% & 85.30\%    & 81.90\%   & 81.60\%    & 81.30\%    & 62.30\%     & 60.80\%     & 41.00\%        & 72.59\% & 88.60\% & 88.20\%    & 83.50\%   & 81.30\%    & 82.10\%    & 68.90\%     & 67.60\%     & 49.70\%        & 76.24\% \\
\multicolumn{1}{c|}{}     & Truncating & 85.60\% & 85.00\%    & 80.70\%   & 80.80\%    & 79.30\%    & 61.00\%     & 59.80\%     & 39.60\%        & 71.48\% & 87.80\% & 87.30\%    & 83.10\%   & 80.70\%    & 81.80\%    & 67.30\%     & 66.90\%     & 47.90\%        & 75.35\% \\
\multicolumn{1}{c|}{}     & Rounding   & 85.90\% & 85.30\%    & 81.50\%   & 81.40\%    & 79.80\%    & 61.50\%     & 60.40\%     & 40.40\%        & 72.03\% & 88.20\% & 87.90\%    & 83.40\%   & 81.30\%    & 81.90\%    & 68.40\%     & 67.40\%     & 49.30\%        & 75.98\% \\
\multicolumn{1}{c|}{}     & Upper      & 85.70\% & 84.90\%    & 80.90\%   & 80.80\%    & 79.40\%    & 61.10\%     & 59.50\%     & 39.40\%        & 71.46\% & 87.80\% & 87.30\%    & 83.00\%   & 80.70\%    & 81.40\%    & 67.00\%     & 66.80\%     & 47.90\%        & 75.24\% \\ \bottomrule
\end{tabular}%
}
\end{table*}

\begin{table*}[htpb]
\centering
\caption{Comparative Analysis of Attack Success Rates using Inception-v4 as Surrogate Model. This table presents a comprehensive evaluation of various attack methods across multiple models.}
\label{aptab:Inc-v4}
\resizebox{\textwidth}{!}{%
\begin{tabular}{@{}cc|ccccccccc|ccccccccc@{}}
\toprule
\multicolumn{2}{c|}{Step}                                    & \multicolumn{9}{c|}{5}                                                                                            & \multicolumn{9}{c}{15}                                                                                            \\ \midrule
\multicolumn{1}{c|}{Attack}                     & Method     & Inc-v3           & Inc-Res-v2           & Res-50           & Res-101          & Res-152          & \begin{tabular}[c]{@{}c@{}}Inc-v3\\ -ens3\end{tabular} & \begin{tabular}[c]{@{}c@{}}Inc-v3\\ -ens4\end{tabular} & \begin{tabular}[c]{@{}c@{}}Inc-Res\\ -v2-ens\end{tabular} & Average          & Inc-v3           & Inc-Res-v2           & Res-50           & Res-101          & Res-152          & \begin{tabular}[c]{@{}c@{}}Inc-v3\\ -ens3\end{tabular} & \begin{tabular}[c]{@{}c@{}}Inc-v3\\ -ens4\end{tabular} & \begin{tabular}[c]{@{}c@{}}Inc-Res\\ -v2-ens\end{tabular} & Average          \\ \midrule
\multicolumn{1}{c|}{\multirow{4}{*}{AdvGAN}}    & DMS        & 65.10\% & 18.40\%    & 60.60\%   & 59.60\%    & 54.80\%    & 33.60\%     & 40.50\%     & 18.60\%        & 43.90\% & 61.50\% & 14.10\%    & 56.20\%   & 52.20\%    & 44.90\%    & 37.10\%     & 32.60\%     & 13.60\%        & 39.03\% \\
\multicolumn{1}{c|}{}                           & Truncating & 60.60\% & 15.60\%    & 57.20\%   & 56.40\%    & 51.50\%    & 30.60\%     & 37.50\%     & 16.00\%        & 40.68\% & 58.40\% & 13.00\%    & 53.30\%   & 49.00\%    & 42.30\%    & 33.40\%     & 29.90\%     & 12.20\%        & 36.44\% \\
\multicolumn{1}{c|}{}                           & Rounding   & 60.40\% & 15.70\%    & 57.00\%   & 56.40\%    & 51.50\%    & 30.60\%     & 37.10\%     & 16.20\%        & 40.61\% & 58.30\% & 13.00\%    & 53.20\%   & 49.40\%    & 42.50\%    & 33.80\%     & 30.00\%     & 12.20\%        & 36.55\% \\
\multicolumn{1}{c|}{}                           & Upper      & 60.40\% & 15.50\%    & 56.80\%   & 56.60\%    & 51.50\%    & 30.60\%     & 37.10\%     & 16.00\%        & 40.56\% & 57.90\% & 13.00\%    & 53.00\%   & 48.90\%    & 42.20\%    & 33.60\%     & 30.00\%     & 12.30\%        & 36.36\% \\ \midrule
\multicolumn{1}{c|}{\multirow{4}{*}{DANAA}}     & DMS        & 90.60\% & 87.20\%    & 86.10\%   & 84.30\%    & 85.10\%    & 64.50\%     & 60.00\%     & 41.90\%        & 74.96\% & 94.50\% & 91.70\%    & 90.60\%   & 90.20\%    & 89.50\%    & 76.70\%     & 73.60\%     & 58.90\%        & 83.21\% \\
\multicolumn{1}{c|}{}                           & Truncating & 89.90\% & 86.10\%    & 85.00\%   & 82.90\%    & 83.80\%    & 62.20\%     & 58.10\%     & 38.80\%        & 73.35\% & 94.30\% & 91.00\%    & 90.10\%   & 89.60\%    & 88.90\%    & 74.50\%     & 71.60\%     & 56.80\%        & 82.10\% \\
\multicolumn{1}{c|}{}                           & Rounding   & 89.80\% & 86.10\%    & 85.10\%   & 83.00\%    & 83.70\%    & 62.10\%     & 58.10\%     & 38.40\%        & 73.29\% & 94.20\% & 90.80\%    & 90.10\%   & 89.70\%    & 88.80\%    & 74.40\%     & 71.10\%     & 56.50\%        & 81.95\% \\
\multicolumn{1}{c|}{}                           & Upper      & 89.80\% & 86.10\%    & 85.20\%   & 82.80\%    & 83.70\%    & 62.10\%     & 58.20\%     & 38.60\%        & 73.31\% & 94.30\% & 90.80\%    & 90.10\%   & 89.70\%    & 88.80\%    & 74.30\%     & 71.40\%     & 56.70\%        & 82.01\% \\ \midrule
\multicolumn{1}{c|}{\multirow{4}{*}{DI-FGSM}}   & DMS        & 61.20\% & 44.60\%    & 41.00\%   & 38.40\%    & 38.70\%    & 16.20\%     & 17.10\%     & 9.50\%         & 33.34\% & 54.60\% & 36.50\%    & 34.70\%   & 31.30\%    & 30.60\%    & 14.20\%     & 15.00\%     & 6.80\%         & 27.96\% \\
\multicolumn{1}{c|}{}                           & Truncating & 59.90\% & 44.00\%    & 41.20\%   & 36.90\%    & 38.60\%    & 16.00\%     & 16.80\%     & 9.60\%         & 32.88\% & 53.60\% & 36.60\%    & 34.20\%   & 31.10\%    & 30.00\%    & 14.10\%     & 14.50\%     & 6.90\%         & 27.63\% \\
\multicolumn{1}{c|}{}                           & Rounding   & 60.50\% & 44.70\%    & 41.40\%   & 38.40\%    & 39.00\%    & 15.80\%     & 17.60\%     & 10.10\%        & 33.44\% & 51.60\% & 32.60\%    & 31.70\%   & 28.60\%    & 28.20\%    & 13.90\%     & 14.50\%     & 6.90\%         & 26.00\% \\
\multicolumn{1}{c|}{}                           & Upper      & 59.60\% & 43.80\%    & 41.00\%   & 37.00\%    & 38.60\%    & 15.90\%     & 16.50\%     & 9.60\%         & 32.75\% & 53.30\% & 36.40\%    & 34.10\%   & 30.80\%    & 29.90\%    & 14.10\%     & 14.50\%     & 7.00\%         & 27.51\% \\ \midrule
\multicolumn{1}{c|}{\multirow{4}{*}{FSPS}}      & DMS        & 88.70\% & 85.80\%    & 83.50\%   & 82.30\%    & 82.50\%    & 75.60\%     & 73.90\%     & 66.20\%        & 79.81\% & 93.60\% & 91.30\%    & 89.40\%   & 87.40\%    & 88.50\%    & 82.90\%     & 81.50\%     & 74.00\%        & 86.08\% \\
\multicolumn{1}{c|}{}                           & Truncating & 88.10\% & 84.70\%    & 83.10\%   & 81.00\%    & 81.60\%    & 74.70\%     & 72.40\%     & 64.40\%        & 78.75\% & 93.20\% & 90.40\%    & 88.50\%   & 86.30\%    & 87.80\%    & 82.40\%     & 80.70\%     & 72.90\%        & 85.28\% \\
\multicolumn{1}{c|}{}                           & Rounding   & 88.20\% & 85.10\%    & 83.40\%   & 81.70\%    & 81.80\%    & 75.20\%     & 72.90\%     & 64.90\%        & 79.15\% & 93.20\% & 90.60\%    & 89.00\%   & 86.70\%    & 88.20\%    & 82.60\%     & 81.20\%     & 73.60\%        & 85.64\% \\
\multicolumn{1}{c|}{}                           & Upper      & 88.10\% & 84.60\%    & 83.00\%   & 81.00\%    & 81.50\%    & 74.70\%     & 72.50\%     & 64.40\%        & 78.73\% & 93.20\% & 90.40\%    & 88.30\%   & 86.30\%    & 87.70\%    & 82.30\%     & 80.50\%     & 72.80\%        & 85.19\% \\ \midrule
\multicolumn{1}{c|}{\multirow{4}{*}{GE-AdvGAN}} & DMS        & 88.40\% & 67.20\%    & 89.10\%   & 91.60\%    & 80.80\%    & 79.70\%     & 80.60\%     & 69.80\%        & 80.90\% & 89.40\% & 69.60\%    & 90.70\%   & 92.80\%    & 84.20\%    & 80.60\%     & 80.50\%     & 72.70\%        & 82.56\% \\
\multicolumn{1}{c|}{}                           & Truncating & 87.50\% & 65.00\%    & 87.40\%   & 91.00\%    & 79.20\%    & 77.40\%     & 78.50\%     & 67.70\%        & 79.21\% & 88.50\% & 66.40\%    & 89.90\%   & 91.30\%    & 81.90\%    & 79.40\%     & 78.60\%     & 71.20\%        & 80.90\% \\
\multicolumn{1}{c|}{}                           & Rounding   & 87.30\% & 65.00\%    & 87.30\%   & 90.70\%    & 79.10\%    & 77.20\%     & 78.30\%     & 67.50\%        & 79.05\% & 88.50\% & 65.80\%    & 90.10\%   & 91.30\%    & 81.30\%    & 79.40\%     & 78.70\%     & 71.20\%        & 80.79\% \\
\multicolumn{1}{c|}{}                           & Upper      & 87.10\% & 65.00\%    & 87.30\%   & 90.80\%    & 79.10\%    & 77.40\%     & 78.50\%     & 67.70\%        & 79.11\% & 88.30\% & 66.00\%    & 90.00\%   & 91.10\%    & 81.70\%    & 79.40\%     & 78.50\%     & 71.20\%        & 80.78\% \\ \midrule
\multicolumn{1}{c|}{\multirow{4}{*}{I-FGSM}}    & DMS        & 47.10\% & 29.90\%    & 30.50\%   & 27.80\%    & 26.50\%    & 11.90\%     & 12.20\%     & 7.20\%         & 24.14\% & 30.70\% & 13.80\%    & 19.00\%   & 17.30\%    & 17.10\%    & 9.10\%      & 11.00\%     & 4.50\%         & 15.31\% \\
\multicolumn{1}{c|}{}                           & Truncating & 45.50\% & 28.80\%    & 30.30\%   & 27.50\%    & 26.40\%    & 11.90\%     & 12.00\%     & 7.10\%         & 23.69\% & 30.10\% & 13.40\%    & 18.40\%   & 17.10\%    & 16.60\%    & 9.90\%      & 10.60\%     & 4.80\%         & 15.11\% \\
\multicolumn{1}{c|}{}                           & Rounding   & 46.60\% & 30.20\%    & 30.80\%   & 28.10\%    & 26.70\%    & 12.00\%     & 12.70\%     & 7.60\%         & 24.34\% & 27.50\% & 12.10\%    & 17.40\%   & 15.40\%    & 14.80\%    & 9.30\%      & 10.70\%     & 4.50\%         & 13.96\% \\
\multicolumn{1}{c|}{}                           & Upper      & 45.40\% & 28.80\%    & 30.30\%   & 27.60\%    & 26.10\%    & 12.00\%     & 12.00\%     & 7.20\%         & 23.68\% & 30.20\% & 13.30\%    & 18.30\%   & 16.90\%    & 16.60\%    & 9.90\%      & 10.80\%     & 4.70\%         & 15.09\% \\ \midrule
\multicolumn{1}{c|}{\multirow{4}{*}{MI-FGSM}}   & DMS        & 63.60\% & 50.40\%    & 50.00\%   & 46.30\%    & 46.20\%    & 19.00\%     & 18.70\%     & 11.10\%        & 38.16\% & 58.80\% & 43.60\%    & 45.20\%   & 41.50\%    & 40.60\%    & 18.90\%     & 18.60\%     & 11.20\%        & 34.80\% \\
\multicolumn{1}{c|}{}                           & Truncating & 63.10\% & 49.90\%    & 49.00\%   & 45.50\%    & 45.30\%    & 18.60\%     & 18.30\%     & 11.00\%        & 37.59\% & 57.90\% & 42.90\%    & 44.40\%   & 40.60\%    & 39.40\%    & 18.10\%     & 18.40\%     & 10.80\%        & 34.06\% \\
\multicolumn{1}{c|}{}                           & Rounding   & 63.00\% & 50.30\%    & 49.00\%   & 45.80\%    & 45.70\%    & 18.90\%     & 18.40\%     & 10.80\%        & 37.74\% & 58.40\% & 43.80\%    & 45.10\%   & 41.20\%    & 40.50\%    & 18.80\%     & 18.90\%     & 11.30\%        & 34.75\% \\
\multicolumn{1}{c|}{}                           & Upper      & 62.90\% & 49.80\%    & 48.70\%   & 45.50\%    & 45.20\%    & 18.50\%     & 18.40\%     & 10.90\%        & 37.49\% & 57.80\% & 43.00\%    & 44.70\%   & 40.80\%    & 39.50\%    & 18.20\%     & 18.20\%     & 10.90\%        & 34.14\% \\ \midrule
\multicolumn{1}{c|}{\multirow{4}{*}{MIG}}       & DMS        & 86.10\% & 79.90\%    & 78.00\%   & 77.20\%    & 77.90\%    & 51.10\%     & 47.40\%     & 33.70\%        & 66.41\% & 84.70\% & 78.40\%    & 77.60\%   & 73.90\%    & 74.60\%    & 51.90\%     & 48.70\%     & 34.10\%        & 65.49\% \\
\multicolumn{1}{c|}{}                           & Truncating & 85.50\% & 78.70\%    & 77.50\%   & 77.00\%    & 77.00\%    & 49.10\%     & 46.20\%     & 32.10\%        & 65.39\% & 84.20\% & 77.80\%    & 76.80\%   & 72.30\%    & 73.50\%    & 50.70\%     & 47.20\%     & 33.60\%        & 64.51\% \\
\multicolumn{1}{c|}{}                           & Rounding   & 85.80\% & 79.20\%    & 78.00\%   & 77.20\%    & 77.30\%    & 49.60\%     & 46.80\%     & 32.70\%        & 65.83\% & 84.60\% & 78.00\%    & 77.40\%   & 73.60\%    & 74.20\%    & 52.30\%     & 48.30\%     & 34.30\%        & 65.34\% \\
\multicolumn{1}{c|}{}                           & Upper      & 85.50\% & 78.70\%    & 77.50\%   & 76.80\%    & 77.00\%    & 49.30\%     & 46.20\%     & 32.00\%        & 65.38\% & 84.20\% & 77.60\%    & 77.10\%   & 72.00\%    & 73.30\%    & 50.80\%     & 47.20\%     & 33.50\%        & 64.46\% \\ \midrule
\multicolumn{1}{c|}{\multirow{4}{*}{NAA}}       & DMS        & 87.80\% & 83.50\%    & 81.20\%   & 81.70\%    & 79.90\%    & 58.90\%     & 54.10\%     & 38.60\%        & 70.71\% & 88.00\% & 84.00\%    & 82.70\%   & 82.70\%    & 81.10\%    & 62.00\%     & 57.80\%     & 41.70\%        & 72.50\% \\
\multicolumn{1}{c|}{}                           & Truncating & 87.10\% & 82.70\%    & 80.70\%   & 81.00\%    & 78.80\%    & 57.00\%     & 52.10\%     & 36.40\%        & 69.48\% & 87.40\% & 83.70\%    & 81.50\%   & 81.80\%    & 80.20\%    & 60.40\%     & 56.70\%     & 39.90\%        & 71.45\% \\
\multicolumn{1}{c|}{}                           & Rounding   & 87.30\% & 82.90\%    & 80.90\%   & 81.20\%    & 79.20\%    & 57.90\%     & 52.50\%     & 37.10\%        & 69.88\% & 87.90\% & 83.80\%    & 82.10\%   & 82.40\%    & 80.80\%    & 61.10\%     & 58.30\%     & 41.50\%        & 72.24\% \\
\multicolumn{1}{c|}{}                           & Upper      & 87.00\% & 82.50\%    & 80.60\%   & 81.00\%    & 78.80\%    & 56.90\%     & 51.80\%     & 36.40\%        & 69.38\% & 87.40\% & 83.60\%    & 81.40\%   & 81.70\%    & 80.10\%    & 60.40\%     & 56.60\%     & 39.80\%        & 71.38\% \\ \midrule
\multicolumn{1}{c|}{\multirow{4}{*}{SINI-FGSM}} & DMS        & 86.70\% & 78.30\%    & 76.50\%   & 75.40\%    & 73.80\%    & 45.00\%     & 42.10\%     & 28.10\%        & 63.24\% & 86.50\% & 78.10\%    & 76.40\%   & 73.50\%    & 72.40\%    & 47.60\%     & 44.30\%     & 29.20\%        & 63.50\% \\
\multicolumn{1}{c|}{}                           & Truncating & 85.80\% & 77.40\%    & 75.20\%   & 73.70\%    & 73.20\%    & 43.70\%     & 41.50\%     & 27.20\%        & 62.21\% & 86.20\% & 77.90\%    & 75.60\%   & 72.40\%    & 72.20\%    & 46.70\%     & 43.60\%     & 28.40\%        & 62.88\% \\
\multicolumn{1}{c|}{}                           & Rounding   & 86.40\% & 78.40\%    & 76.00\%   & 74.90\%    & 73.60\%    & 44.10\%     & 42.00\%     & 27.30\%        & 62.84\% & 86.70\% & 78.10\%    & 76.20\%   & 73.10\%    & 72.90\%    & 47.70\%     & 44.30\%     & 29.20\%        & 63.53\% \\
\multicolumn{1}{c|}{}                           & Upper      & 85.70\% & 77.10\%    & 75.20\%   & 73.50\%    & 72.80\%    & 43.90\%     & 41.50\%     & 27.20\%        & 62.11\% & 85.90\% & 77.70\%    & 75.50\%   & 72.40\%    & 72.00\%    & 46.70\%     & 43.60\%     & 28.40\%        & 62.78\% \\ \midrule
\multicolumn{1}{c|}{\multirow{4}{*}{SSA}}       & DMS        & 89.40\% & 85.00\%    & 81.30\%   & 79.30\%    & 79.70\%    & 70.80\%     & 69.50\%     & 58.40\%        & 76.68\% & 91.60\% & 88.00\%    & 83.90\%   & 81.60\%    & 82.80\%    & 77.10\%     & 75.80\%     & 65.90\%        & 80.84\% \\
\multicolumn{1}{c|}{}                           & Truncating & 88.80\% & 84.50\%    & 80.40\%   & 77.90\%    & 78.60\%    & 68.90\%     & 68.40\%     & 57.00\%        & 75.56\% & 91.20\% & 87.50\%    & 82.80\%   & 81.20\%    & 82.10\%    & 75.60\%     & 74.90\%     & 64.70\%        & 80.00\% \\
\multicolumn{1}{c|}{}                           & Rounding   & 89.30\% & 84.70\%    & 81.30\%   & 78.60\%    & 79.10\%    & 70.00\%     & 68.70\%     & 57.50\%        & 76.15\% & 91.30\% & 87.90\%    & 83.60\%   & 81.40\%    & 82.60\%    & 76.60\%     & 75.50\%     & 65.40\%        & 80.54\% \\
\multicolumn{1}{c|}{}                           & Upper      & 88.80\% & 84.40\%    & 80.20\%   & 77.70\%    & 78.80\%    & 68.90\%     & 68.20\%     & 57.10\%        & 75.51\% & 91.20\% & 87.50\%    & 82.80\%   & 80.80\%    & 82.20\%    & 75.40\%     & 74.50\%     & 64.70\%        & 79.89\% \\ \midrule
\multicolumn{1}{c|}{\multirow{4}{*}{GRA}} & DMS        & 88.40\% & 83.30\%    & 80.70\%   & 79.80\%    & 80.30\%    & 63.40\%     & 60.10\%     & 45.60\%        & 72.70\% & 90.70\% & 86.00\%    & 82.60\%   & 81.80\%    & 82.40\%    & 70.60\%     & 69.10\%     & 56.50\%        & 77.46\% \\
\multicolumn{1}{c|}{}     & Truncating & 87.50\% & 82.80\%    & 79.50\%   & 78.80\%    & 79.50\%    & 62.50\%     & 58.80\%     & 44.50\%        & 71.74\% & 90.60\% & 85.40\%    & 82.00\%   & 81.20\%    & 81.40\%    & 69.40\%     & 67.70\%     & 55.50\%        & 76.65\% \\
\multicolumn{1}{c|}{}     & Rounding   & 87.70\% & 83.10\%    & 80.30\%   & 79.10\%    & 79.80\%    & 62.80\%     & 59.10\%     & 45.10\%        & 72.13\% & 90.60\% & 86.00\%    & 82.70\%   & 81.70\%    & 82.20\%    & 70.50\%     & 68.80\%     & 56.40\%        & 77.36\% \\
\multicolumn{1}{c|}{}     & Upper      & 87.50\% & 82.80\%    & 79.30\%   & 78.70\%    & 79.40\%    & 62.30\%     & 58.70\%     & 44.50\%        & 71.65\% & 90.60\% & 85.30\%    & 81.90\%   & 80.70\%    & 81.10\%    & 69.30\%     & 67.70\%     & 55.70\%        & 76.54\% \\ \bottomrule
\end{tabular}%
}
\end{table*}

\end{document}